  

 \magnification=\magstep1
\settabs 18 \columns
\hsize=16truecm

\input epsf

\def\b{\bigskip}
\def\bb{\bigskip\bigskip}

\def\no{\noindent}
\def\r{\rightline}
\def\ce{\centerline}
\def\ve{\vfill\eject}

\def\r{\rightline}

\def\L{{\cal L}}

\def\harr#1#2{\smash{\mathop{\hbox to .25 in{\rightarrowfill}}
 \limits^{\scriptstyle#1}_{\scriptstyle#2}}}

\def\today{\ifcase\month\or January\or February\or March\or April\or
May\or June\or July\or
August\or September\or October\or November\or  December\fi
\space\number\day, \number\year }

\r \today
\bb\bb\bb

\def\DD{\vec \bigtriangledown}

\def\w{\wedge}

\def\p{\partial}

\def\sqr#1#2{{\vcenter{\vbox{\hrule height.#2pt
\hbox{\vrule width.#2pt height#2pt \kern#2pt
\vrule width.#2pt}
\hrule height.#2pt}}}}

 \def\1/2{{\scriptstyle{1\over 2}}}
 \def\a/2{{\scriptstyle{3\over 2}}}
 \def\5/2{{\scriptstyle{5\over 2}}}
 \def\7/2{{\scriptstyle{7\over 2}}}
 \def\3/4{{\scriptstyle{3\over 4}}}



 

 \def\1/2{{\scriptstyle{1\over 2}}}
 \def\a/2{{\scriptstyle{3\over 2}}}
 \def\5/2{{\scriptstyle{5\over 2}}}
 \def\7/2{{\scriptstyle{7\over 2}}}
 \def\3/4{{\scriptstyle{3\over 4}}}

\font\steptwo=cmb10 scaled\magstep2

\magnification=\magstep1

\vskip2cm
\vskip-3cm\ce{\steptwo  A classical mistake and what it tells us.} 

\ce{\bf How to do better with an action principle}

\ce{\bf for Hydro and Thermo
dynamics}
\bb

\ce {Christian Fronsdal}

\ce{\it Dept. of Physics and Astronomy, University of California Los Angeles, USA}
\bb\bb 

{\it ABSTRACT} ~~~ Rayleigh's stability analysis of cylindrical Couette flow, of 1889 and 1916, is in contradiction with observation. The
analysis is repeated in many textbooks and reviews up to 2017, and its failure to agree with observation was duly noted. More successful approaches have been found, but little was done to discover the weak point of Rayleigh's argument,
what is the reason that it fails.  This paper identifies the mistake as one that is endemic in the literature. Since the physics of the problem remains poorly understood, a discussion
of this paradox should prove useful.
Briefly, the argument depends on the Navier-Stokes equation and on the assumption that a certain expression  called  ``energy density" or
``kinetic potential" can be interpreted and used  as  such. It is shown here that 
the use of any expression as a kinetic potential is in conflict with the Navier-Stokes equation, in all but a very limited context.   

An alternative analysis of basic Couette flow, based  on an  action principle for compressible fluids, provides a Hamiltonian density as well as a kinetic potential. The two are not the same, even in the simplest cases.  
The action principle provides a kinetic potential; a new criterion for stability recognizes the profound effect of the surface adhesion and the tensile strength of water. It is in full agreement with observation. 
  Several new experiments are suggested.

\bb


\ve

\no{\bf I. Introduction}

An action principle due to Lagrange (1760, 1781), rediscovered by Lamb (1932) and by Fetter and Walecka (1980),  gives  us an efficient and elegant formulation of
hydrodynamics.  The Galilei invariant action in modern notation is
$$
A_{FW} = \int dt\int d^3x\left(\rho(\dot\Phi -\DD\Phi^2/2 - \phi - W[\rho]\right).\eqno(1.1)
$$ 
The velocity field is irrotational, 
$$
\vec v = -\DD\Phi,\eqno(1.2)
$$
which is a strong limitation. There have been attempts to lift the action principle to a more general context, especially in dealing with superfluids, and some have come very close to formulating a more general action principle. The strong hint provided in a widely cited paper (Hall and Vinen 1956) was overlooked.  

Traditional hydrodynamics has 4 independent variables, the density and the 3 
components of the velocity. The action (1.1) has only 2 independent variables,
the density and the velocity potential; two more are needed. But it will not do to seek a direct generalization of Eq.(1.2).

The scalar velocity potential is essential, for several reasons. In the first place, the Euler-Lagrange equation
$$
{\delta \over\delta \Phi}A_{FW} =  \dot\rho +\DD\cdot (\rho\vec v)=0\eqno(1.3)
$$
is the equation of continuity, the very essence of hydrodynamics.

As important is the tradition that incorporates gravitation into field theories, especially into hydrodynamics,  by including  the Newtonian potential in the Hamiltonian density.  (Granted that it is suggested by the correspondence between hydrodynamics and particle mechanics, but our subject is field theories.) Newtonian hydrodynamics is a non-relativistic approximation to General Relativity, but  is there a generally-relativistic field theory that has the expected non-relativistic limit?  Yes, there is. 

 Consider the generally relativistic field theory with action (Fronsdal 2007)
 $$
 \int d^4x\sqrt{-g}\left({\rho\over 2}(g^{\mu\nu}\psi_{,\mu}\psi_{,\nu} -c^2)
 -W[\rho]\right).\eqno(1.4)
 $$
Gravitation is represented by the metric, in the non relativistic limit by the time-time component. To explore the  non-relativistic limit set
 $$
 g_{00} = c^2-2\varphi, ~~~g_{11} = g_{22} = g_{33} = -1, 
 $$
 other components zero, expand (1.4) in powers of (1/c) and take the 
 limit $1/c\rightarrow 0$. To ensure the cancellation of the terms of
lowest order  we need to set
$$
\psi = c^2t + \Phi, 
$$
then terms of order $c^2$ cancel and we are left with the action (1.1).   In the context of field theories this is the best (perhaps the only)   confirmation of the expectation that  Non-Relativistic Hydrodynamics, with the inclusion of the gravitational potential, is the non-relativistic limit of General Relativity.  Without the expansion $\psi = c^2t + \Phi$ the role of gravitation  in hydrodynamics would loose the only direct contact that it has with General Relativity. Equation (1.3) leads to the correct transformation law for the field $\Phi$ under Galilei transformations, given that $\psi$ is a scalar field on space time.    
But this theory is limited to irrotational flows.

The equations of hydrodynamics are Galilei invariant, but the 
Hamiltonian is not; only the action is invariant; the search for a Galilei invariant expression for the energy is mis-directed. See {\it e.g.} 
Khalatnikov (1955  page 57), Putterman (1974 page 22).

We conclude that the required generalization of  (1.1) must include the field
$\Phi$ and  additional fields with 2 degrees of freedom, with an action that reduces to (1.1) by projection. Two additional dynamical variables are needed, one additional pair of canonical variables, either an additional scalar field or else a gauge field with only one propagating component.

The direct introduction of another scalar velocity potential would not 
accomodate rotational flow, and the only appropriate relativistic gauge theory, with a single propagating mode, is the theory of the massless 2-form field $(Y_{\mu\nu})$ of Ogievetskij and Polubarinov (1964), the theory of the notoph.  

The connection of potential flow to an action principle is intimately related to the fact that the velocity is a gradient of a more basic field, a scalar potential. The velocity field of an alternative, `Lagrangian' formulation of hydrodynamics, usually denoted $\dot{\vec x}$, is also a derivative and also a natural candidate for a canonical variable. We shall call this vector potential $\vec X$,  to distinguish it from the coordinates. Since the  equations of motion are second order in the time derivative this field appears to have 6 independent degrees of freedom. We already have 2 degrees of freedom with $\rho$ and $\Phi$, conventional hydrodynamics has 4, so this would be far too much. It turns out, however, that it is possible to impose a constraint that reduces the number of degrees of freedom of the new field to 2. The constraint leads to the  non-relativistic limit of the relativistic notoph gauge theory, with a single propagating mode.  The complete Lagrangian density is, in a fixed gauge,
$$
\L = \rho(\dot\Phi +\dot{\vec X}^2/2 +\kappa\rho \dot{\vec X}\cdot\DD\Phi - \vec\Phi^2/2 - \phi) - W[\rho].\eqno(1.5)
$$
The aptness of this Lagrangian will be confirmed in the next section, by  comparison with Navier-Stokes theory. 

The full, relativistic gauge theory is well known (Ogievetskij and Palubarinov 1964).  
\b

\ce{\bf Constraint and  Euler-Lagrange equations}

Constraints are typical of gauge theories and arise from variation of the action with respect to a gauge field (components $Y_{0i}$ of a 4-dimensional  2-form), in this case the constraint is
$$
\DD\w \vec m = 0,~~~\vec m := \rho\vec w,~~~\vec w := 
\dot{\vec X} + \kappa\DD\Phi).\eqno(1.6)
$$
This is the constraint that reduces the number of additional degrees of freedom carried by the field $\vec X$ to 2. The gauge is fixed by setting the gauge field to zero and only the coponents $X^i\epsilon_{ijk} = Y_{jk}$ appear in our gauge-fixed, non-relativistic Lagrangian. The Euler-Lagrange equations are as follows.

Variation of the velocity potential $\Phi$ gives the equation of continuity,
$$
\dot\rho +\DD\cdot\rho\vec v,~~~\vec v := \kappa\dot{\vec X}  - \DD\Phi.\eqno(1.7)
$$
The flow vector is thus $\rho\vec v$, with $\vec v$ defined in (1.7). Variation of $\vec X$ leads to
$$
{d\over dt}\vec m = 0.\eqno(1.8)
$$
The field $\vec m$ that appears in (1.6) and in (1.8) has been called `the momentum' (Lund and Regge 1976), it is distinct from $\rho\vec v$. Finally, variation of the density gives, when the force of gravity is neglected, the
integrated Bernoulli equation (Bernoulli 1738) 
$$
\dot\Phi + \dot{\vec X}^2/2 +\kappa\rho \dot{\vec X}\cdot\DD\Phi - \DD\Phi^2/2   = \mu.\eqno(1.9)
$$  

The significance of the parameter $\kappa$ will become clear as we study the applications.  The action of the Galilei group on the notoph field is a gauge transformation; the gauge-fixed field $\vec X$ is inert.

The  proposed Lagrangian, Eq.(1.5), has much in common with modern theories of rotational flow; they all involve potential flow plus additional velocity fields (Fetter 2009). From Eq.s(1.1-4) it is seen that the vorticity field is 
$$
\DD\w\vec v = \kappa\DD\w\vec w  = \kappa\DD{1\over \rho}\w\vec m = (\DD{\kappa\over \rho})\w \vec w;\eqno(1.10)
$$
it is orthogonal to the density gradient:
$
\DD\rho \cdot  (\DD\w\vec v) = 0. 
$
Incompressible fluids can be understood only as a limit within a theory of compressible fluids. Supporting this statement is the fact that there can be no
incompressible fluids in a relativistic theory.

\bb

\no{\bf II.  Cylindrical Couette flow}

Couette flow has been been studied for 130 years, with the focus on the spontaneous onset of turbulence. The experiment has a
homogeneous  fluid contained in the space between two concentric cylinders,
as in Fig.(2.1), both cylinders turning independently. A completely  satisfactory treatment has yet to be found (Lin 1955, Betchov and Criminale 1967,  Joseph 1976,
Drazin and Ried 1981, Schmid and Henningson 2000). 
 

A pioneering series of experiments, by Couette (1887-1890) and Mallock (1888)
was followed by an important paper by G.I. Taylor (1923). 
This work included new experiments as well as a penetrating analysis, the main impact of which was the vindication of the `non-slip' boundary condition. It is the statement that the limit of the material flow velocity, at the  boundary of the containing vessel, is the same as the local velocity of the container at the same point.
(Rayleigh, writing in 1916,  was already using the non-slip boundary condition 
without comment, in the proof of the theorem under discussion.)
   \vskip1.5in  
   
\vskip-3cm
 
\epsfxsize.4\hsize
\centerline{\epsfbox{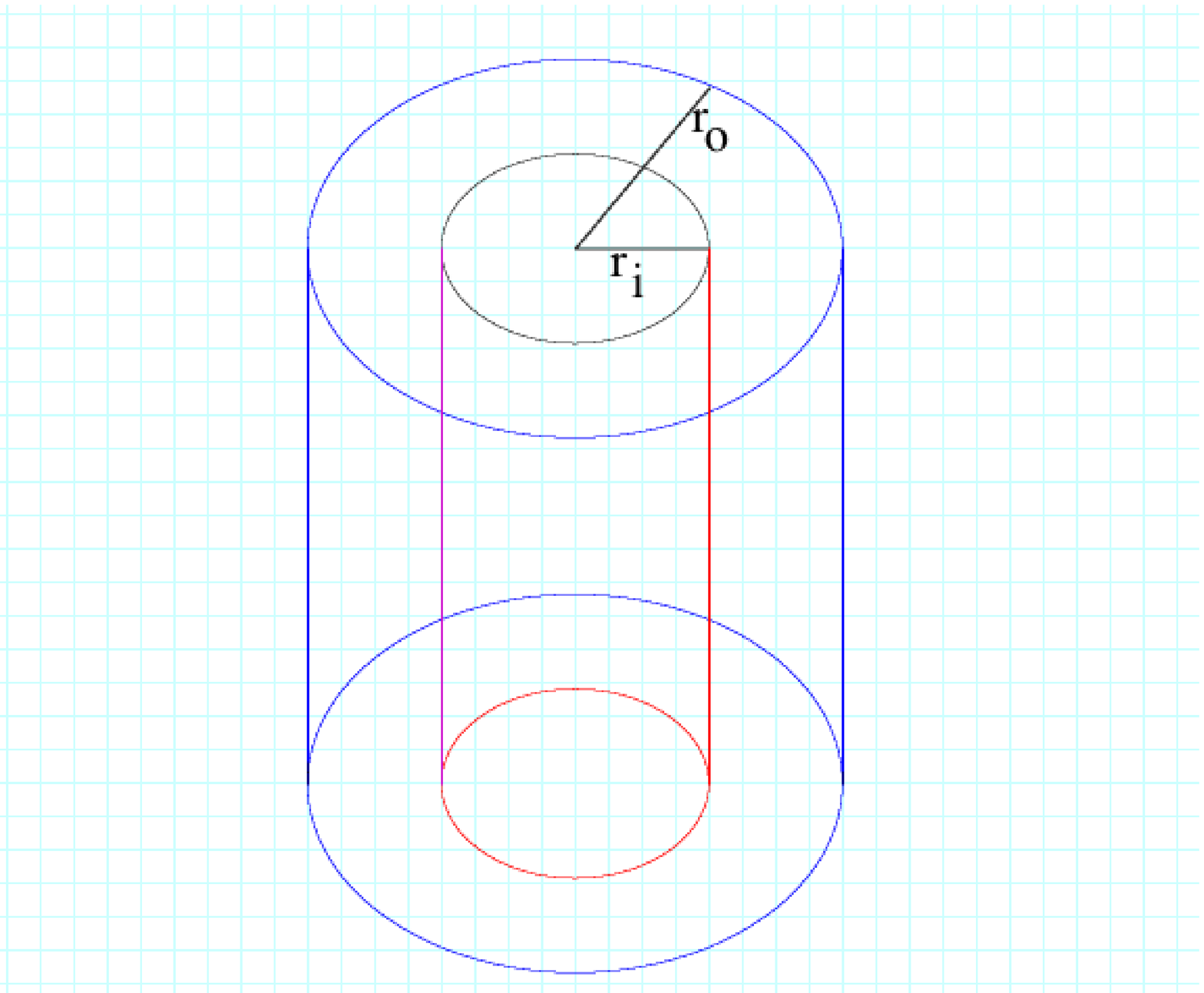}}
\vskip-.5cm

\parindent=1pc
\bb

Fig.2.1. Cylindrical Couette flow, both cylinders turning independently.

\vskip-2cm
 
\epsfxsize.4\hsize
\vskip-.5cm

\parindent=1pc

\bb
   \vskip2in  

\vskip-3cm
 
\epsfxsize.4\hsize
\centerline{\epsfbox{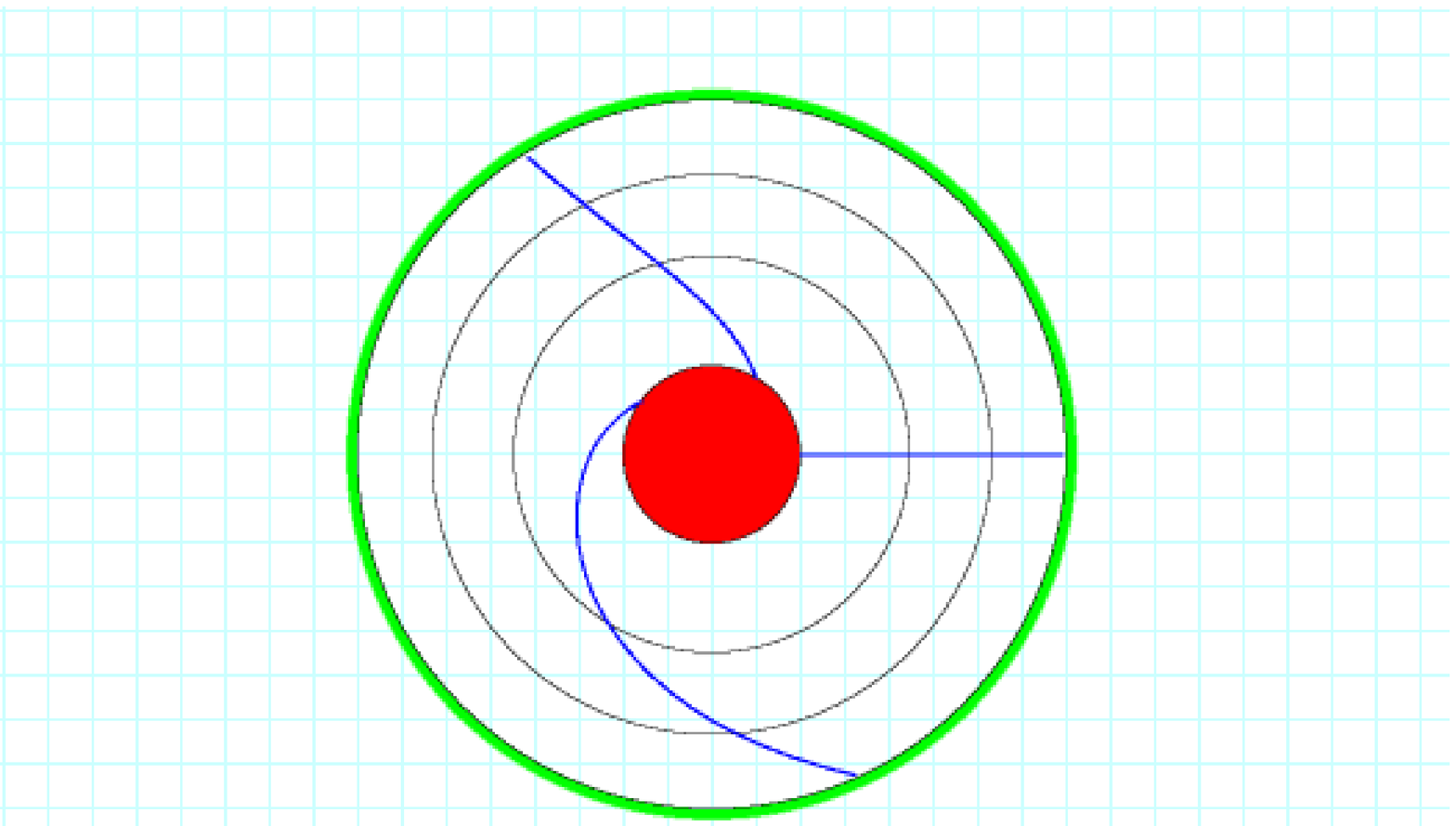}}
\vskip-.5cm

\parindent=1pc
\bb

Fig.2.2. Illustration of Cylindrical Couette flow. The curved lines are 
successive snap shots of a set of particles originally in the radial configuration.

\b

Experiments start with both cylinders at rest, then they are rotated,
independently, with  
angular speeds $\omega_i,\omega_o$  that are increased slowly, until instability of the flow is observed. The most interesting presentations of experimental results have been in the form of a partition of the plane of angular speeds into a stable and an unstable region. The borderline, similar to a hyperbola, marks the onset of instability, the stable region lies below.  See Fig. 5.2.
\b

\ce{\bf Solid body  flow}

The simplest type of flow velocity is  stationary, horizontal and circular, in cylindrical coordinates,$$
\dot{\vec X} = \omega(r)(-y,x,0), ~~~~r :=\sqrt{x^2+y^2}.  
$$
The boundary conditions are non-slip. To simplify the analysis one thinks of the cylinders as being long, and ignore end effects.

The particular case
$$
\dot{\vec X} = b(-y,x,0), ~~~~b~~{\rm constant},\eqno((2.1)
$$
is the flow of a solid body. It gives rise to a kinetic potential
$$
K = -\dot{\vec X}^2 \propto -r^2,\eqno(2.2) 
$$
(all proportionalities with positive coefficients) and a centrifugal force 
$
-\DD K \propto \vec r
$
 that is balanced by a pressure (constant + $r^2$)  that presents a force - $\DD p \propto -\vec r$. 
  It follows, for ordinary fluids with a positive  adiabatic derivative $d p/ d\rho$, if there are no other forces acting, that $K+p$ is stationary and that the density  must increase outwards. See Fig 2.1.
\bb
 
\epsfxsize.3\hsize 
\centerline{\epsfbox{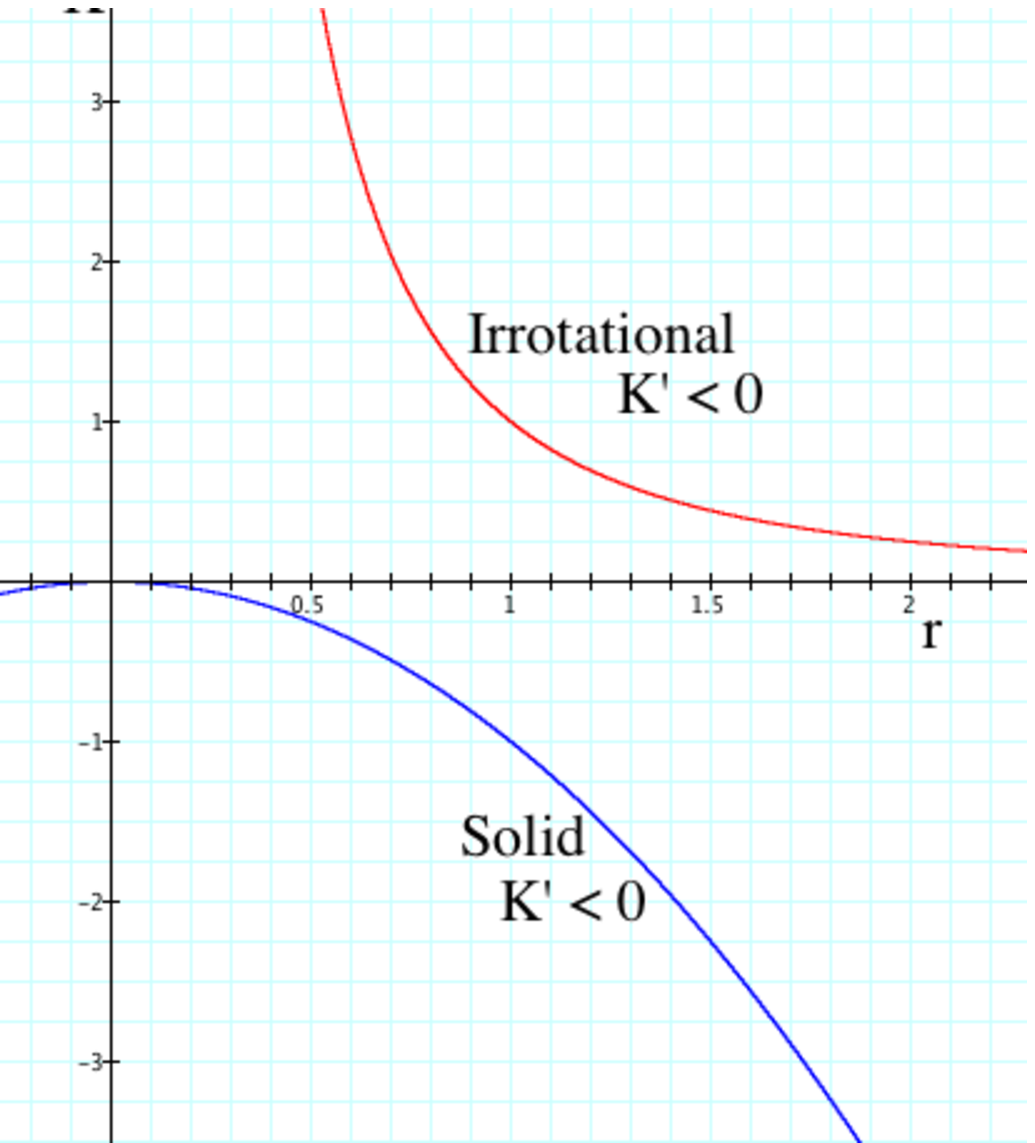}}

\parindent=1pc

\vskip1cm
Fig.2.1.  The kinetic potentials, $1/r^2$ (irrotational flow) and $-r^2$ (solid-body flow).
\b 

The  kinetic potential is $\vec v^2/2$ for irrotational flow and $-\vec v^2/2$
for solid-body flow. 
Disturbing question: what is the expression for the  `kinetic potential'
when the velocity has 2 parts, one rotational and one  solid-body type?
 \b 

\ce{\bf The Navier-Stokes equation}

\no is the basis for every traditional analysis of Couette flow: 
$$
{D\over Dt} \vec u = -{1\over \rho}\DD p + \bar\mu \rho\Delta\vec u.\eqno(2.3)
$$
Here $\bar\mu$ is the kinematical viscosity of the fluid, in the simplest case a constant parameter. This equation agrees with the gradient of (1.9) in the special case that $\vec u = -\DD\Phi$ and $\dot{\vec X}$ = 0. It also agrees with (1.9) in the complimentary case, when $\DD\Phi = 0$ and $\vec u =  \dot{\vec X}$. (The velocity field in (2.3) is denoted $\vec u$ to emphasize that it should not be prematurely identified with either $-\DD\Phi$ or $\dot{\vec X}$ or $\vec v$.)

Stationary flow is possible only if the effect of viscosity is negligible,
this requires that $\bar\mu = 0$ or else that 
$$
\Delta \vec u = 0.\eqno(2.4)
$$

When the flow is stationary, and in the planes perpendicular to the axis, as observed at low speeds, then this condition  allows the general horizontal flow
$$
\vec u = {a\over r^2}(-y,x,0) + b(-y,x,0),~~~a,b ~~{\rm constant},\eqno(2.5)
$$
the parameters $a,b$ to be determined by the non-slip boundary conditions.
The first is irrotational for $r>0$,
$$
{a\over r^2}(-y,x,0) = a\DD \theta,~~~ \theta = \arctan{y\over x}.
$$
This gives rise to a kinematical potential 
$$
K\propto \DD\Phi^2\propto 1/r^2\eqno(2.6)
$$
 and a centrifugal force 
 $$
 -\DD K \propto \vec r/r^4,
 $$
and in this case too the force points outwards.
Both of these types of flow are of great interest, especially so in connection with superfluid Helium.   The second term in (2.5) is the flow of a solid body,
see Eq.s (2.1), (2.2).  We observe that $dK/dr$ is negative in both cases
and that $K$ is proportional to $u^2$ in one case and to $-u^2$ in the other (all
proportionalities with positive coefficients). 

These elementary considerations already suggest that there are difficulties
ahead.   What is the kinetic potential when the velocity
field is of neither type, but a combination of both, as in (2.5)?

This prediction of the Navier - Stokes equation is a brilliant success, being in very good agreement with experiments, at low speeds of rotation. The only difficulty is that there is no clear guide to constructing  an expression for the energy density, or a kinetic potential. It would be natural to expect that the expression  
$$
\rho\vec u^2/2,~~~{\rm resp.} ~~~ \vec u^2/2,
$$
might be an important part of it, serving as a kinetic potential and giving rise to the centrifugal acceleration; we have seen that this not the case in general, though it is true in the case of irrotational flow,  as we shall now show, again.
\b

\ce{\bf The question of sign, again}

Consider the simplest cases. The total derivative in (2.3) is
$$
{D\over Dt}\vec u := {d\over dt}\vec u + (\vec u\cdot\DD )\vec u.
$$
the last term is (minus) the centrifugal force. When the flow is of the solid body type, Eq.(2.2),
$$
 (\vec u\cdot\DD )\vec u = b^2(-y\p_x+xd_y)(-y,x,0) = -b^2(x,y,0) 
= -\DD\vec u^2/2.
$$
That is, the acceleration is outwards and the centrifugal potential 
 $K$ is $- \vec u^2/2$. In the complimentary case of potential flow,
$$
 (\vec u\cdot\DD )\vec u = {a^2\over r^2}(-y\p_x+x\p_y){1\over r^2}(-y,x,0) 
=  -{a^2\over r^4}(x,y,0)  = \DD(\vec u^2/2); 
$$
it is in the same direction, hence correct, but now the centrifugal potential is
 $\vec u^2/2$.  The difference in sign between the two cases means that 
\b

\ce {\bf One can not, in general, associate the Navier Stokes equation} 

\ce{ \bf with an energy density  $E[\rho, \vec u]$ or with a kinetic potential $K[\vec u]$.}
\b

This conclusion is long overdue.

\ve

\ce{\bf  A classical mistake and what it tells us} 


 To theorists the principal aim has been to understand the limits on the stability of the laminar flow that was described above and the first notable attempt  to do so  was that of Rayleigh (1889, 1916) who concluded that the laminar flow should be stable if and only if 
  $$
 \omega_o > \omega_i~~~(\rm approximately).
 $$
 Here $\omega_o (\omega_i)$ is the angular velocity of the outer (inner)
 cylinder and  $\omega_i$ is positive by convention. In particular, this implies  that the laminar flow would be unstable whenever the two cylinders rotate in opposite directions. The result of observation was different: the condition for stable Couette flow is, approximately, 
$$
\omega_o^2 > \omega_i^2,
$$
which allows for stable flow for either sign of $\omega_o/\omega_i$.

\b

Raleigh's  reaches his  conclusion twice. 
\b

1. Angular momentum (about the axis of rotation) is conserved, the  angular momentum is $L = vr$,   ``... so that the centrifugal force acting on a {\it given portion} of the fluid is  $L^2/r^3$". That is,  $ - L^2/r^2$ is regarded as a kinetic potential. But this is without justification and, as it happens,
actually wrong; if there is any kinetic potential $K$ in the context of the Navier-Stokes equation it must  satisfy  
$$
(\vec u\cdot \DD)\vec u = \DD K.\eqno(2.7)
$$
2.  We quote (Rayleigh 1916):  ``We may also found our argument upon a direct consideration  of the kinetic energy of the motion.....". He concludes that stability requires that this function increase outwards. This is the same argument and here we see more clearly what the difficulty is: There is no  justifiable way to introduce the energy concept, or to choose an expression to serve as energy,  except in the context of an action principle.

\b

Since the Navier - Stokes equation is an expression of the balance of forces
 one must ask if there is an additional force that must be taken into account.
Yes, there is; but what is without justification is the claim that there is an `energy density' proportional to $L^2/r^2$ and that it functions as a kinetic potential.

 We have seen (1) that any valid expression for `kinetic potential' must have opposite signs for the two types of flow  and that (2) if there is a suitable expression for a kinetic potential in the context of the Navier-Stokes equation, then it is not the energy density.
 
  Rayleigh's choice of `energy' leads to conclusions that are contradicted by experiment. If we merely restrict ourselves to the two special cases examined in the preceding section we see that there is, in either case, an effective kinetic potential, proportional to $\vec u^2$, but with a positive coefficient in one case and a negative coefficient in the other; this disqualifies it from being interpreted as an energy in the general case and  blocks Rayleigh's argument.

During the 100 years that followed the publication of Rayleigh's paper the calculation has been repeated in numerous textbooks, including these: 
Chandrasekhar (1955 and 1980), Landau and Lifshitz (1959),  Drazin and  Ried (1981), Tilley and Tilley (1986), Koshmieder (1993) and  Wikipedia (2017).

That Rayleigh's prediction was contradicted by experiments must have been known to himself in 1916; that it was known to the other authors mentioned is not in doubt. Yet there is no  suggestion in the later literature  that Rayleigh's argument is faulty!

Landau and Lifshitz (1959, page 100) claim that viscosity makes Rayleigh's argument invalid if the cylinders rotate in opposite directions! Chandrasekhar (1981, page 275) argues that the criterion is inconsequential because the region in which it is violated is small, Drazin and Ried (1981, page 79) repeat Chandrasekhar's argument. Koschmieder ({\it op cit} Chapter 11) offers the most comprehensive discussion
of traditional methods  but he too finds no fault with Rayleigh's argument ({\it op cit} Chapter 10). Finally, Wikipedia repeats Rayleigh's argument but offers no clue to why the criterion fails when the cylinders are rotating in opposite directions.

The Navier-Stokes equation by itself does not imply, and in general it does not allow, the existence of an expression with the attributes of energy.  But a
kinetic potential can be constructed in the special case of stationary, laminar Couette flow, when  the velocity field is of the form (2.5); it is (Fronsdal 2014)
$$
-K =  \ {b^2\over 2}r^2 +  ab \ln r^2  -{a^2\over 2r^2},~~~~
(\vec u\cdot\DD)\vec u = \DD K.\eqno(2.8)
$$
By the Navier-Stokes equation, this makes a contribution $-\DD K$ to the acceleration.  Therefore,this
function is the only kinetic potential that is consistent with the Navier-Stokes equation in the simplest case when the velocity is of the form (2.5). It is not  ($1/\rho$ times the kinetic part of) the Hamiltonian.
\footnote * { In  the highly regarded book {\it Stellar Structure and Evolution"}, by Kippenhahn, Weigert and Weiss (212), we find definitions of normalized gravitational, internal and total energies. 
Then they  take it for granted that the  name given to the expression for the `total energy' endows it with  physical properties. In this manner is reproduced a correct result previously proved by Chandrasekhar (1935).   Other examples of such shortcuts are legion, See e.g. Eddington (1926, page 142). Here is different type of example. Arnold and Khesin in their book 
(1998, pages 2, 19, 37, 75, 119) lay down axioms that, for them, define Hydrodynamics. They always use the standard definition of `energy', $E= v^2/2$.}

\ve

\no{\bf III. The action principle includes a kinetic potential}

 The Hamiltonian density is
$$
h = \rho\big(\dot {\vec X}^2/2 + (\DD\Phi)^2/2 + \phi\big) + f+sT.\eqno(3.1)
$$ 
Instead of the expected  square of the `total velocity' we have the sum of two squared velocities. But the equation  of motion, obtained by taking the gradient of (1.9), is
$$
\DD\dot\Phi - \DD\left(-\dot{\vec X}^2/2 -\kappa\dot{\vec X}\cdot\DD\Phi + (\DD\Phi)^2/2 + \phi\right) = {1\over \rho}\DD p, \eqno(3.2)
  $$
where $p$ is the thermodynamic pressure.
 Equations (3.1-2) both have the correct signs, the first gives the Hamiltonian density with the correct, positive  sign for both  terms, the second equation agrees with the Navier-Stokes equation (when applicable) with the two different signs. The expression in the large parenthesis is a kinetic potential, but it is not simply related to an energy density.
 
The Navier - Stokes equation combines the two types of flow in a single velocity field and deals correctly with both of them, but that is as far as one can go with a single velocity field, for the irrotational flow is Eulerian and the solid-body flow is Lagrangian.

 
 We conclude that the action principle is not in conflict with the traditional treatment of this type of flow, but it completes it by giving us the equation of continuity, as well as an energy density; the Hamiltonian is a first integral of the motion and the Lagrangian, not the Hamiltonian, contains the  kinetic potential. 
  
With a Hamiltonian and a kinetic potential in hand we can apply standard methods.  The first  result is, of course, that a stable configuration must satisfy the 
 Euler-Lagrange equations, making the energy stationary  with respect to all perturbations. A deeper analysis studies harmonic perturbations to first and
 second order of perturbation theory, as in Chandrasekhar (1955 and 1980), Landau and Lifshitz (1959),  Drazin and  Ried (1981) and  Koshmieder (1993).   
 
 In the presence of viscosity there can be no energy conservation and no action principle. But there is a standard and  natural way to modify the Euler-Lagrange equations. Instead of (2.3) one poses
 $$
 {d\over dt} \vec m = \bar\mu\rho \Delta \vec v.\eqno(3.3)
 $$
We have limited our attention to stationary flows, with $\Delta \vec v = 0$,
taking our inspiration from the traditional point of view.

\b

\ve

\no{\bf IV.  Stability of laminar Couette flow by the action principle}


According to (3.2), the kinetic potential is the function
$$
K =  -\dot{\vec X}^2/2 - \kappa\rho \dot{\vec X}\cdot\DD\Phi + \DD\Phi^2/2.\eqno(4.1)
$$

By the gauge constraint, for horizontal, circular flow there are constants $a,b$ such that
$$
- \DD\Phi = {a\over r^2}(-y,x,0),~~~ \vec m = -\DD \tau = {b\kappa\over r^2}(-y,x,0).\eqno(4.2)
$$
The second formula solves the gauge constraint (1.6).                                                   The velocity of mass transport is 
$$
\vec v = \kappa{\vec m\over \rho} - (1+\kappa^2)\DD\Phi = 
\left({\kappa^2 b\over r^2\rho} + (1+\kappa^2) {a\over r^2}\right)(-y,x,0)
= {1+\kappa^2\over r^2}({c^2b\over\rho} + a)(-y,x,0),\eqno(4.3) 
$$
with $c^2 :=\kappa^2/(\kappa^2+1)$. The only constraint on the field $\vec v$ is that it must be harmonic - see Eq.3.3; thus
$$
{1\over \rho} =  1 + \alpha (1-r^2),~~~\alpha~{\rm constant}.\eqno(4.4)
$$
This density profile implies that the vorticity - Eq. (1.10) - is uniform, value $2\alpha\kappa b$.

We have referred to the classical intuitive feeling that the pressure must increase outwards. For any normal fluid, for which the adiabatic derivative $dp/d\rho$ is positive, this implies that the density must increase outwards, so the constant $\alpha$ must be positive.
 (We have normalized the density at the outer boundary to unity.)  This relates $\alpha$ to $\rho_i$,
$$
\alpha = 4.54(1/\rho_i - 1).\eqno(4.5)
$$

In terms of these vector fields  
$$
-K = {\vec m^2\over 2\rho^2} - (\kappa^2+1)\DD\Phi^2/2=
 {1+\kappa^2\over 2r^2}\left({b^2c^2\over \rho^2} - a^2\right),~~~
c^2 :={\kappa^2\over 1+\kappa^2}.\eqno(4.6)
$$
By (4.5),  
$$
-K = {1+\kappa^2\over 2}\left( b^2c^2\big({(1+\alpha)^2\over r^2} -2\alpha(\alpha+1) +  \alpha^2r^2\big)-{a^2\over 2r^2}\right)
\eqno(4.7)
$$
and
$$
-K' = {1+\kappa^2\over r}\bigg({a^2\over r^2} + b^2c^2(\alpha^2r^2-{(1+\alpha)^2\over r^2})\bigg).\eqno(4.8) 
$$
\b

Now that we have a kinetic potential we can apply Rayleigh's criterion, 
$K' < 0$. 

\ve 
 \b

\epsfxsize.5\hsize
\centerline{\epsfbox{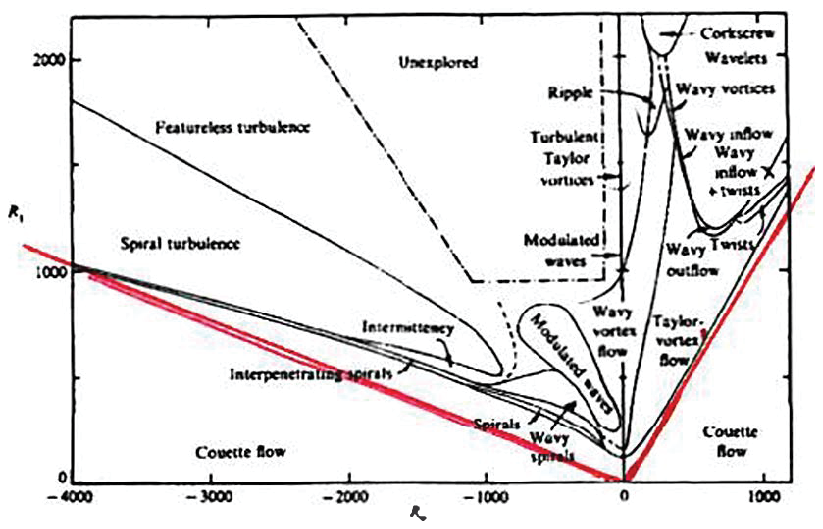}}
\
Fig.4.1. A pair of straight through the origin is the best possible fit of a locus $K'=0$ to the ``hyperbolic" curve that has been observed.
 \b

\b

\ce{\bf Boundary conditions}

The walls of the two cylinders move with angular  velocities

$$
\omega_i\hat\theta = {\omega_i\over r}(-y,x,0),~~ \omega_o\hat\theta
 = {\omega_o\over r}(-y,x,0)~~~\omega_i,  \omega_o~~{\rm constant}. \eqno(4.9)
$$
The velocity of mass transport is $\vec v$, Eq.(4.3), this is the velocity that must satisfy  the no-slip boundary conditions, whence
$$
{1+\kappa^2\over r_i^2}(a+{bc^2\over\rho_i}) = \omega_i,
~~~{1+\kappa^2\over r_0^2}(a+{bc^2\over\rho_o}) = \omega_o,\eqno(4.10)
$$
and
$$ 
a = {r_o^2\rho_o\omega_o - r_i^2\rho_i\omega_i\over (\rho_o-\rho_i)(\kappa^2+1)},
~~~~ b = \rho_o\rho_i{r_i^2\omega_i - r_o^2\omega_o\over \kappa^2(\rho_o-\rho_i)}.\eqno(4.11)
$$

\b
When these quantities are expressed in terms of the angular velocities
we see that the locus $K' = 0$ consists of 2 straight lines through the origin of the angular velocity diagram. The best result 
that can be obtained with the assumption that $K'$ must be negative, is for the lines $K'=0$ to coincide with the `asymptotes' of the experimental curve, as in Fig. 4.1. That would leave a substantial region that is stable, although $K'$ is positive. \

But even that is not be possible.

The zero locus of (4.8)
consists of the two lines
$$
{a\over bc} = \pm \beta,~~~ \beta = \sqrt{(1+\alpha)^2-\alpha^2r^4},
$$
or
$$
{\omega_i\over \omega_o} = {1\pm c/\beta r^2 \rho_i\over 1\pm c/\beta r^2}.\eqno(4.12)
$$
If the two asymptotes are in the first and fourth quadrant it follows that either
$$
 ~~c< \beta < c/\rho_i,~~~{\rm or~ else}~~~  c > \beta > c/\rho_i.
\eqno $$
  In the first case $\rho_i<1$. Then $\alpha>0$. At the outer boundary 
$$
\beta^2 = 1+2\alpha = 1+2{1/\rho_i-1\over 1- r_i^2} < 1/\rho_i^2.
$$
By (4.12), $\beta^2 =   1+2\alpha < c^2/\rho_i^2 < 1/\rho_i^2$,
 which reduces to $2/(1- r_i^2) < 1+1/\rho_i$,  which is false since the density is very close to 1. At the inner surface, $\beta$ is even larger.
 In the second case $\rho_i > \rho_o$, which is anti-intuitive.

This suggests that the criterion $K'<0$ cannot be made to account for observations, for any choice of the parameters. Later we shall find that
that this condition is satisfied almost everywhere; it is not sufficient
and it does not relate to the observed boundary of stability, except in a very limited sense. 
\bb

\no{\bf V. A proposal}

It is known that some types of  instability are accompanied by bubble formation. (As in  the wake of propellers.) We suggest that this may be related to local evacuation (or more generally to a physical breakdown of the laminar nature of the flow) and
that it happens at a particular value of the chemical potential, the density and  of the corresponding value the kinetic potential $K$. 

We shall test this hypothesis. We are not rejecting the idea that $K'$
must be negative; it turns out that it is  negative almost everywhere. 
 
\b

Let us subtract the constant term from $K$ in (4.7).
As we see from that equation, the locus of $K=0$ consists of two straight lines.  At the inner boundary, 
$$
{\omega_i\over \omega_0} = ({r_o\over r_i})^2{1\pm c\over 1\pm c(\rho_i/\rho_o)}\eqno(5.3)
$$
and these must have opposite signs, hence
$
1 > c >  {\rho_o\over \rho_i}.
$
At the outer boundary,

$$
{\omega_i\over \omega_o} = ({r_o\over r_i})^2{1\pm c(\rho_o/\rho_i)\over 1\pm c)}
$$
and $ 1>c>\rho_i/\rho_o$.
This implies that, as the velocity is increased, the instability first manifests itself at the boundary with the higher density. 

The value of $K$ varies with the radius $r$. If, as the velocity is increased, the instability  makes a first appearance at the  boundary $B$, the inner or the outer cylinder, then we expect that the parameters can be tweaked to make a locus $K(r_i) = A$ or $K(r_o)=A$   an adequate approximation to the experimental limit curve and that the  family of loci, $K(r) = A, r_i < r < r_o$  displace upwards as we move away
from the boundary  $B$. 

 We first examined the possibility that the instability first manifests itself  at the outer boundary, which would imply that the density is higher there, as had been expected. This resulted in a failure; we shall say no more about it. 
Then we examined the alternative, that the instability first manifests itself  at the inner boundary,  in which case the density is higher there, against expectations. This calculation  will now be described in detail. 
\bb

\ce{\bf Main  results}

Assume that
$
\rho_i > \rho_o.
$
Fig. 5.1, taken from Andereck {\it et al}  (1883, 1986) is a summary of their results, the lowest line, resembling a hyperbola, is the upper limit of observed, basic Couette flow.
 
Fig 5.2 shows a $K$ locus that best fits the experimental curve, with
 $\rho_o/\rho_i = .9, \kappa = 5.3$ and $R=r_i$. Identical curves are obtained with $\rho_o/\rho_i = .99, ~\kappa = 17.5$, $\rho_o/\rho_i = .999, ~\kappa = 55$ and so on, calculated  up to $\rho_o/\rho_i = 1-10^{-6}$.
   The slope of the right asymptote is predicted without ambiguity to be very close to $(r_o/r_i)^2$, -- a bullseye for the theory. See Eq. (5.3). The slope on the left is very sensitive to the value of   $\kappa$; this too can be seen by inspection of (5.3).

\vskip0
cm 
   
\b
      
\vskip-2mm

\epsfxsize.5
\hsize
\centerline{\epsfbox{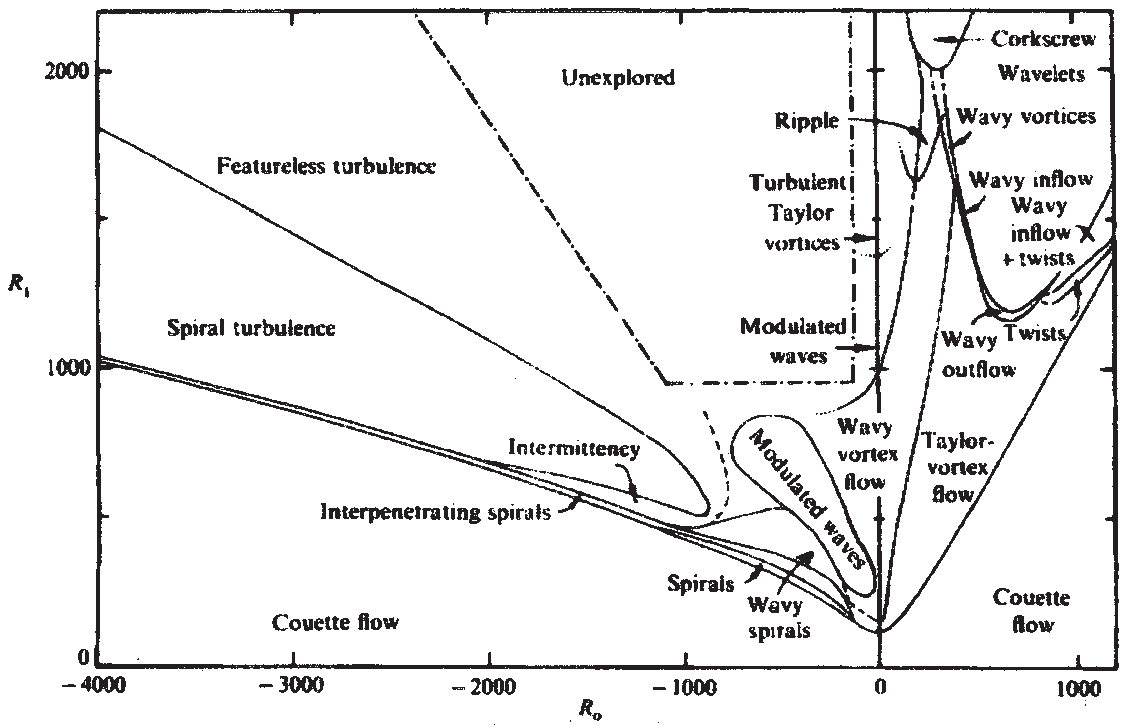}}
\vskip-.5cm

\parindent=1pc

\bb

Fig. 5.1. Experimental results of Andereck et al (1983, 1986). The abscissa (ordinate) is the angular velocity of the outer (inner) boundary. The lower `hyperbola' is the upper limit of stability of laminar Couette flow.  
\b

\vskip0cm

\epsfxsize.5
\hsize
\centerline{\epsfbox{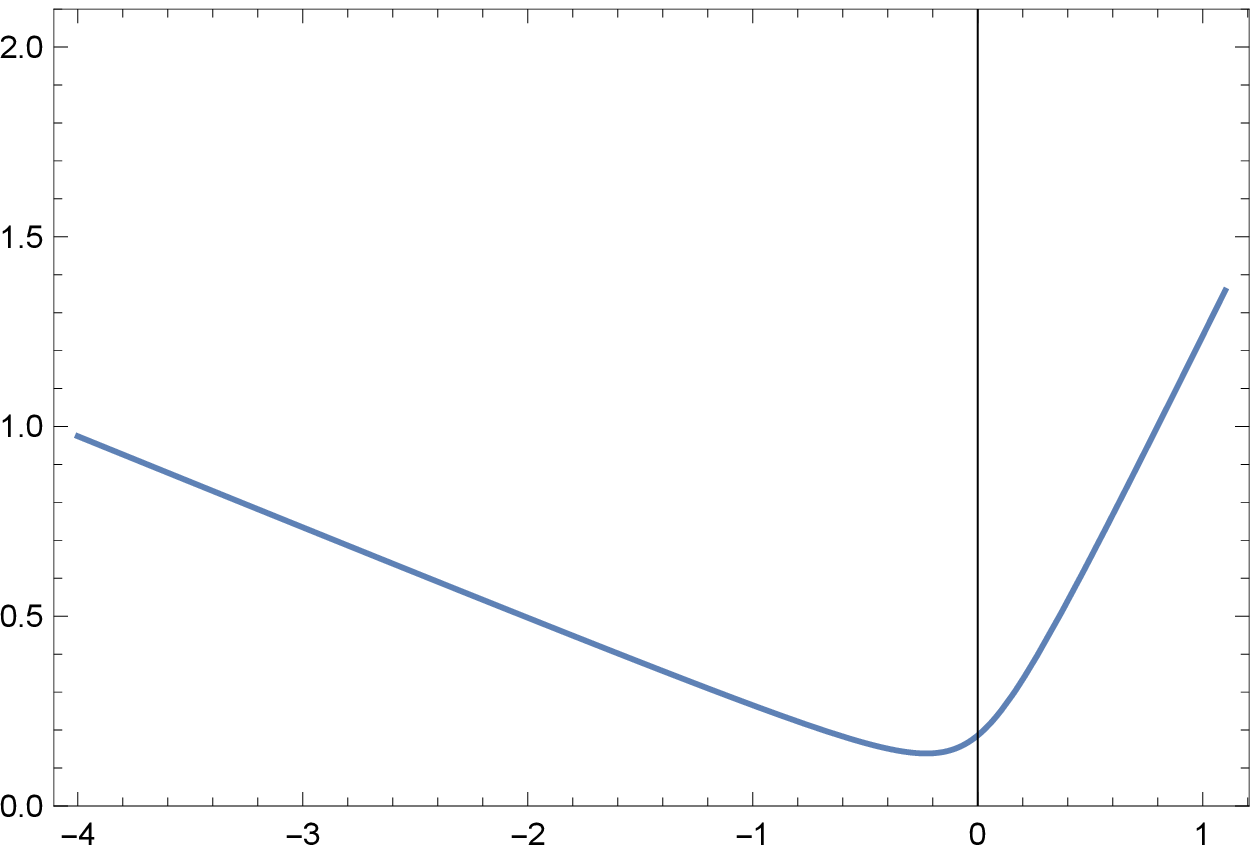}}
\vskip0cm

\parindent=1pc
 
Fig.5.2. A locus $K(r_i) = A_+ > 0$ = constant. The lines $K=0$ lie below.
 \ve

The subsequent figures show the result of repeating the calculation, with the same values of the parameters and the same value of $A$, at several points in the interval $r_i <r< r_o$.
\vskip.5cm

\epsfxsize.5\hsize
\centerline{\epsfbox{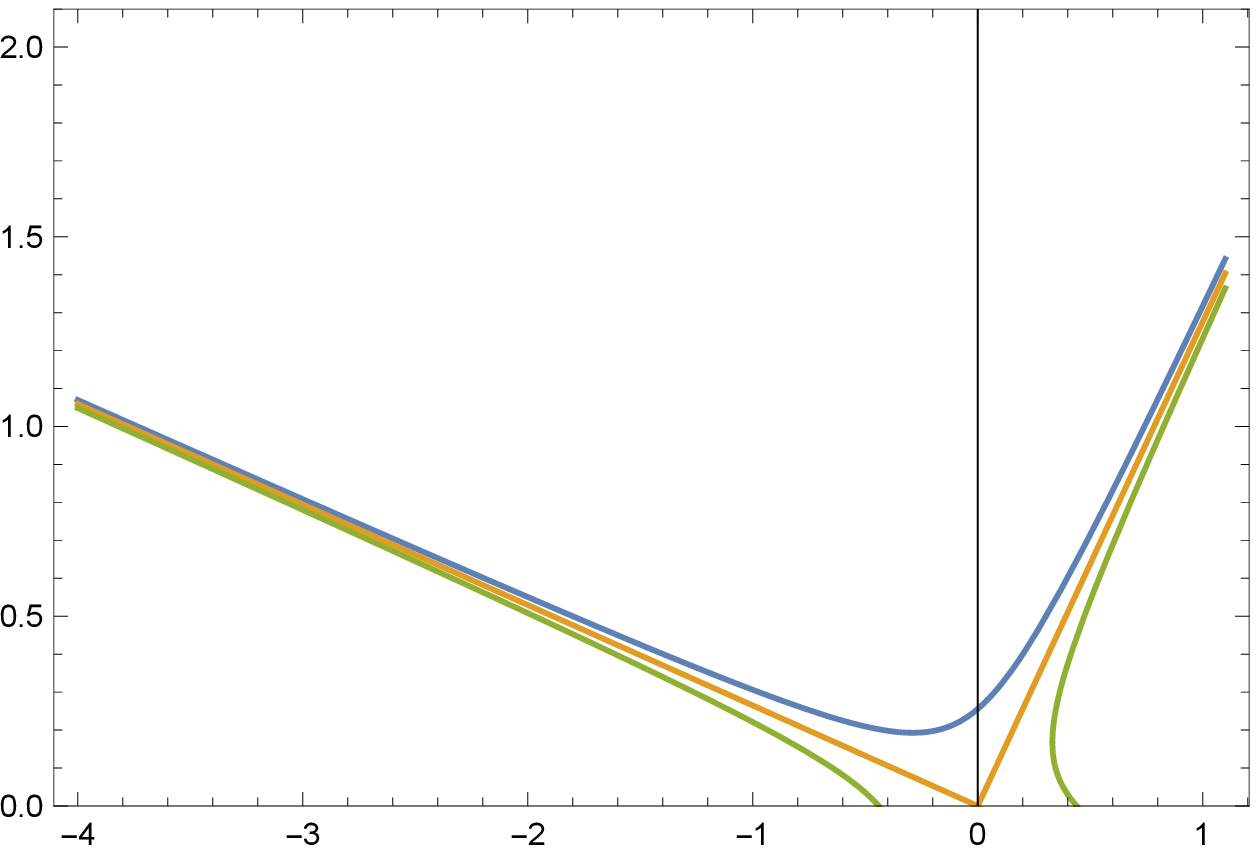}}

\vskip0cm

\vskip-.8cm
 
 Fig.5.3. Loci $K=A_- <0, 0, A_+>0$ at the inner boundary, $r = r_i = .883$. The three curves are drawn for the same values of $K$ in all the diagrams that follow.
\vskip.5cm
\epsfxsize.5\hsize
\centerline{\epsfbox{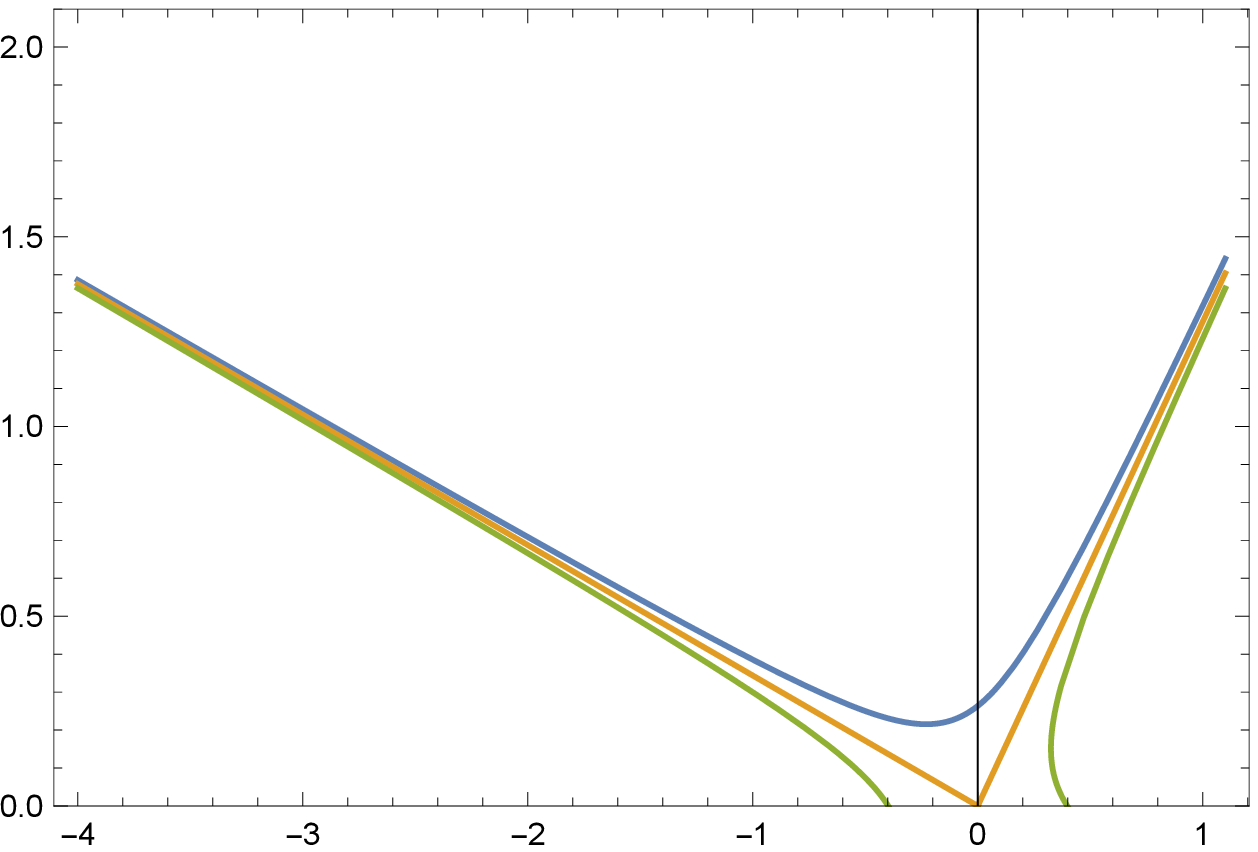}}
\vskip0cm 

\vskip-1cm
 
 Fig.5.4. The locus $K=A$ with the same values of $A$ at $r=.888$.
 \vskip.5cm
\epsfxsize.5\hsize
\centerline{\epsfbox{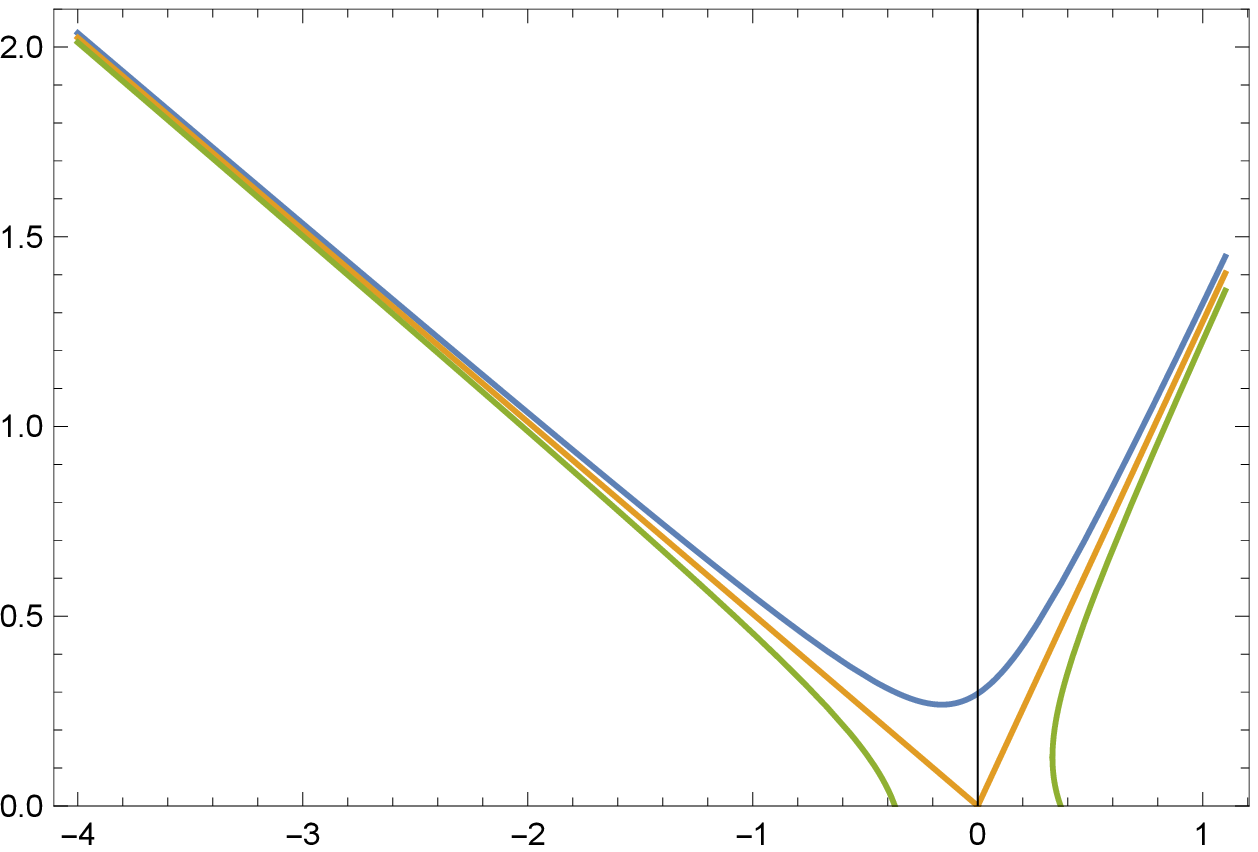}}
\vskip-.7cm

Fig.5.5. At $r = .893$. the locus is moving upwards, almost everywhere.

\vskip0cm
 
\epsfxsize.5\hsize
\centerline{\epsfbox{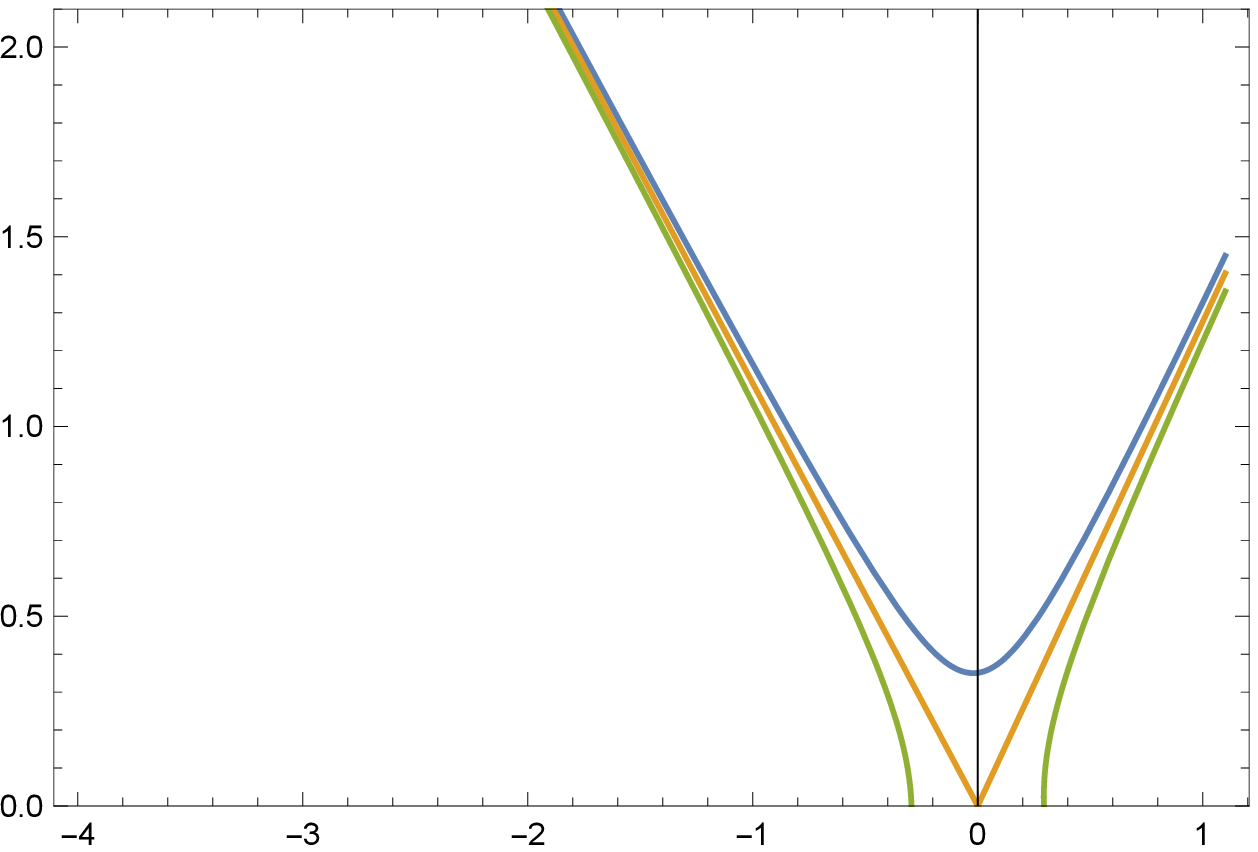}}
\vskip0cm

Fig.5.6. At $r = .898$.
\vskip.5cm
 
\epsfxsize.5\hsize
\centerline{\epsfbox{Mistake92.eps}}
\vskip0cm

Fig.5.7. At $r = .92$.

\epsfxsize.5\hsize
\centerline{\epsfbox{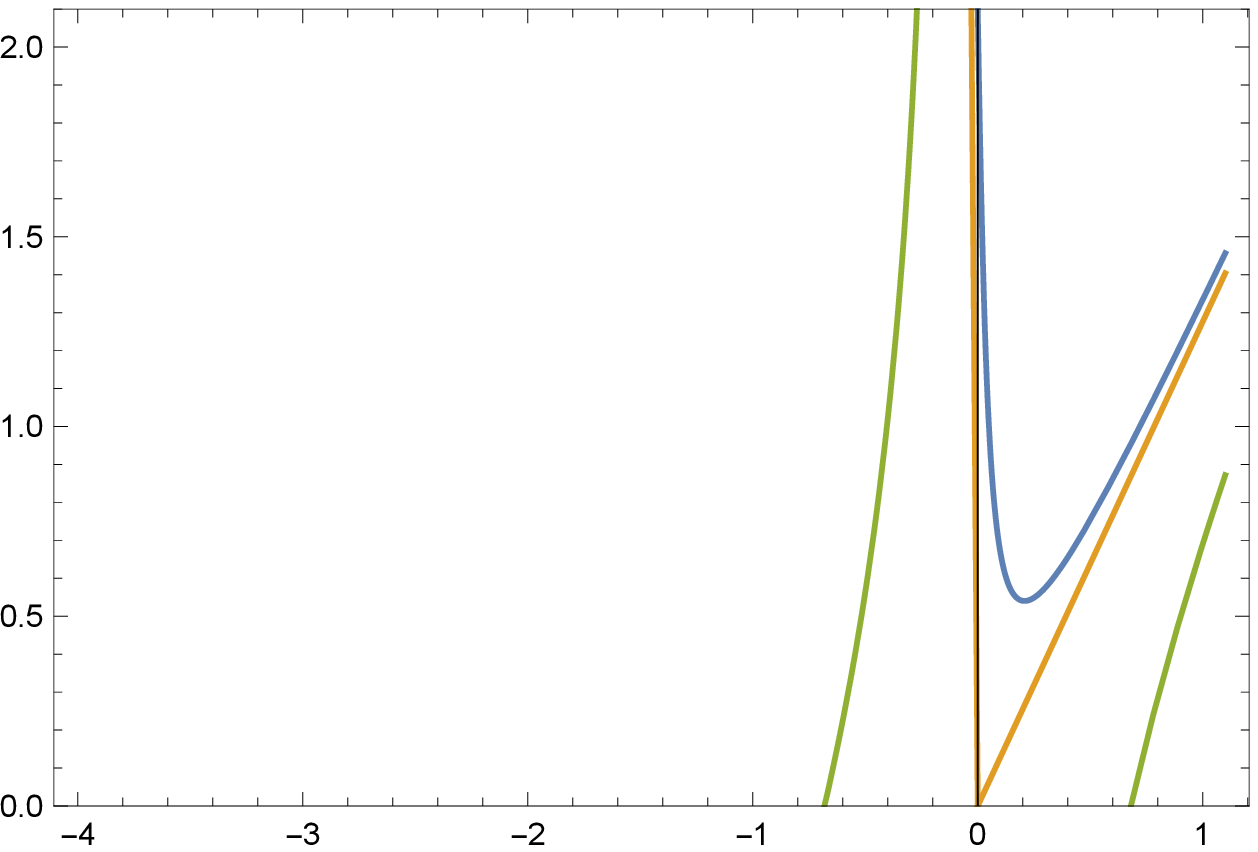}}
\vskip0cm

Fig.5.8. And at $r=.98$.

\b 

In each of  these diagrams two straight lines through the origin make up the $K=0$ locus. The hyperbola that lies between them is a locus  $K=A_+>0$. The two lines that lie outside form a locus  $K = A_- < 0 $. All these figures have the same parameters and the same values of  $A_+$ and $A_-$. The series start at the inner boundary and proceeds to the outer boundary.

The crucial question  is  what happens to the locus $K(r)=A$ as $r$
is increased from $r = r_i$ towards $r = r_o$, for fixed values of the parameters and for a fixed value of $A$.
Comparing  Figures 5.5-10 it can be seen that, almost everywhere, the locus moves upwards with increasing radius; hence at any point in the diagram, the value of $K$ decreases.  
Only the region near the right asymptotes is difficult to assess.
Diagrams 5.3 - 5.8 show that the domain of stability gradually expands as
we move outwards, $K'$ is negative almost everywhere and this function does not provide a criterion for stability.

To get a different view of the situation we next plot a zero locus of $K'$, with the same parameters. As the result is virtually independent of the radius in the range $r_i < r < r_o$; we show only the case $r=r_i$. It is seen that $K'$ is negative everywhere except in a small region on the right, which confirms what we concluded by visual inspection of the $K$ loci. 
\b

\epsfxsize.5
\hsize
\centerline{\epsfbox{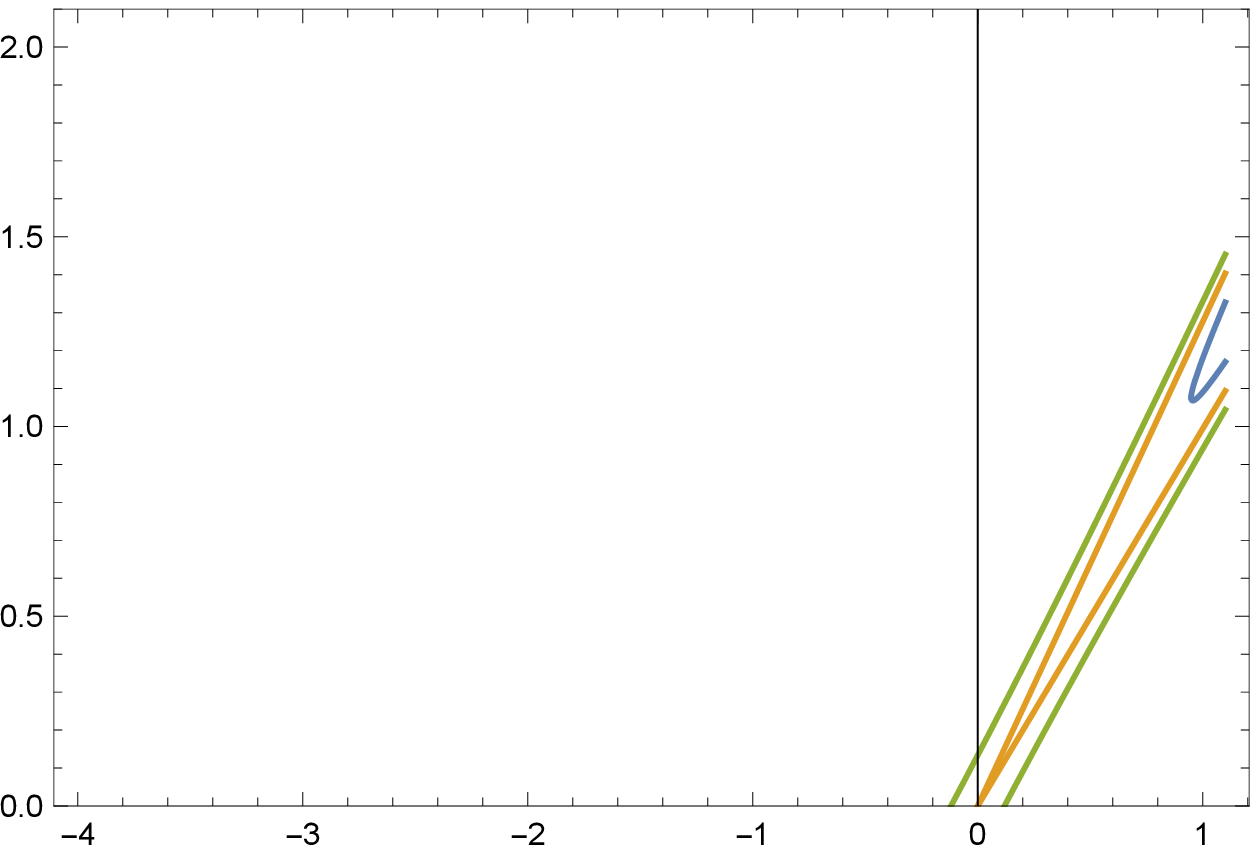}}
\vskip0cm

\parindent=1pc
\b

Fig. 5.9. These are loci of the derivative of the the kinetic potential. It is zero on the two straight lines through the origin and it is positive between them, everywhere else it is negative.
\b 

That is reassuring, except for the fact that the pressure gradient  has the `wrong' sign since, to balance it,  the centrifugal force it should have the wrong sign as well. Instead of two forces balancing each other, the centrifugal force and the pressure  gradient reinforce each other!

The next two illustrations are flow lines, in this sense. If at some instant
we drop a line of saw dust on the horizontal straight line that connects the two cylinders on the right side (on the positive ``x"-axis), then this is what we shall see at some future instant. One gets the impression that the volume elements are being dragged along by the ones closest to the walls, and perhaps  this suggests an increased pressure near the inner cylinder.

If the  behavior of the system seems to be contrary to our intuition, then one thing that can be done is to educate our intuition.

Are we forgetting a third force? Yes, and  perhaps a fourth.

\ve

\vskip1cm

\epsfxsize.5
\hsize
\hskip-3cm
\centerline{\epsfbox{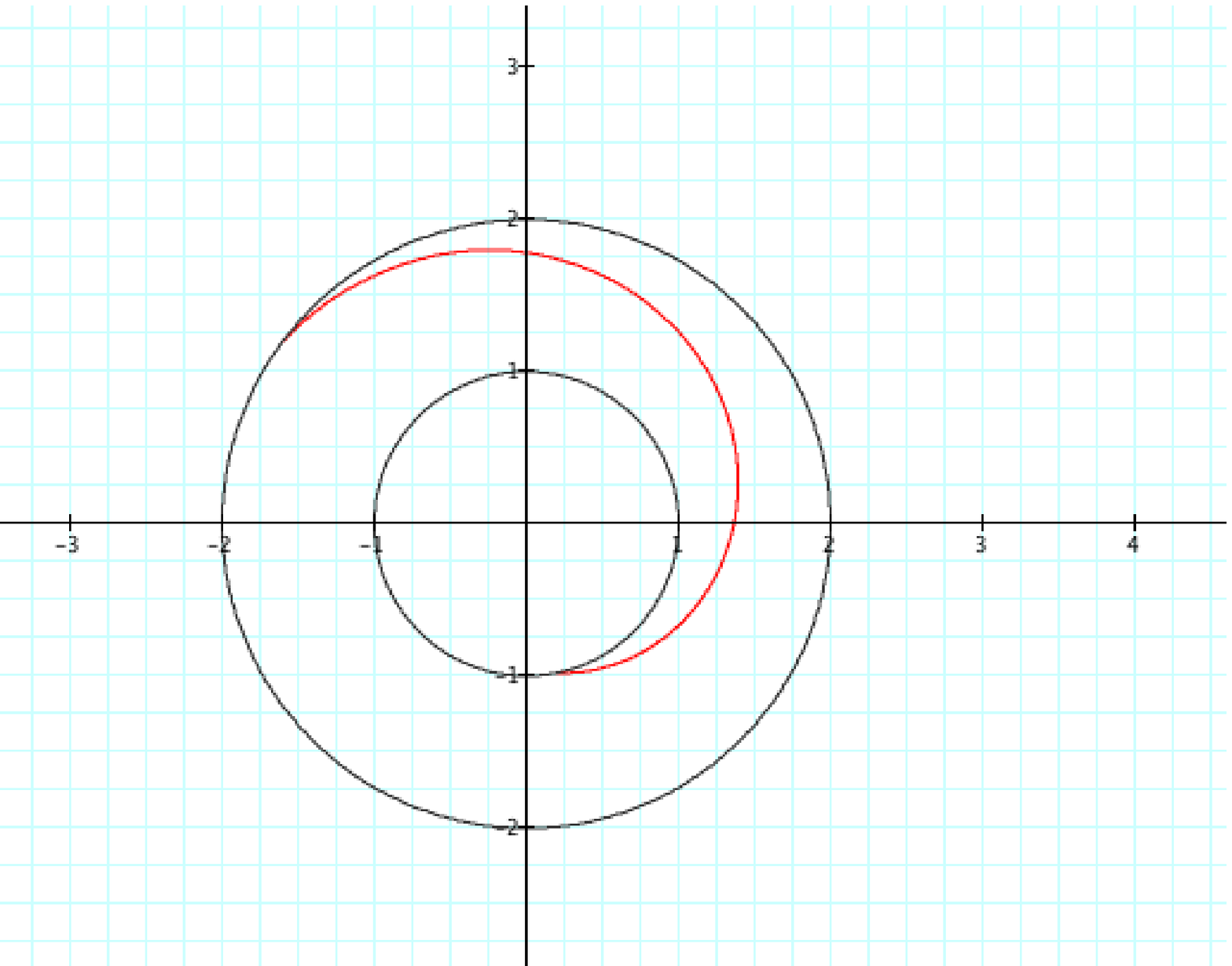}}
\vskip0cm

\parindent=1pc\hskip1mm
\b

\vskip-2.3in\hskip1.5in
\epsfxsize.4\hsize
\centerline{\epsfbox{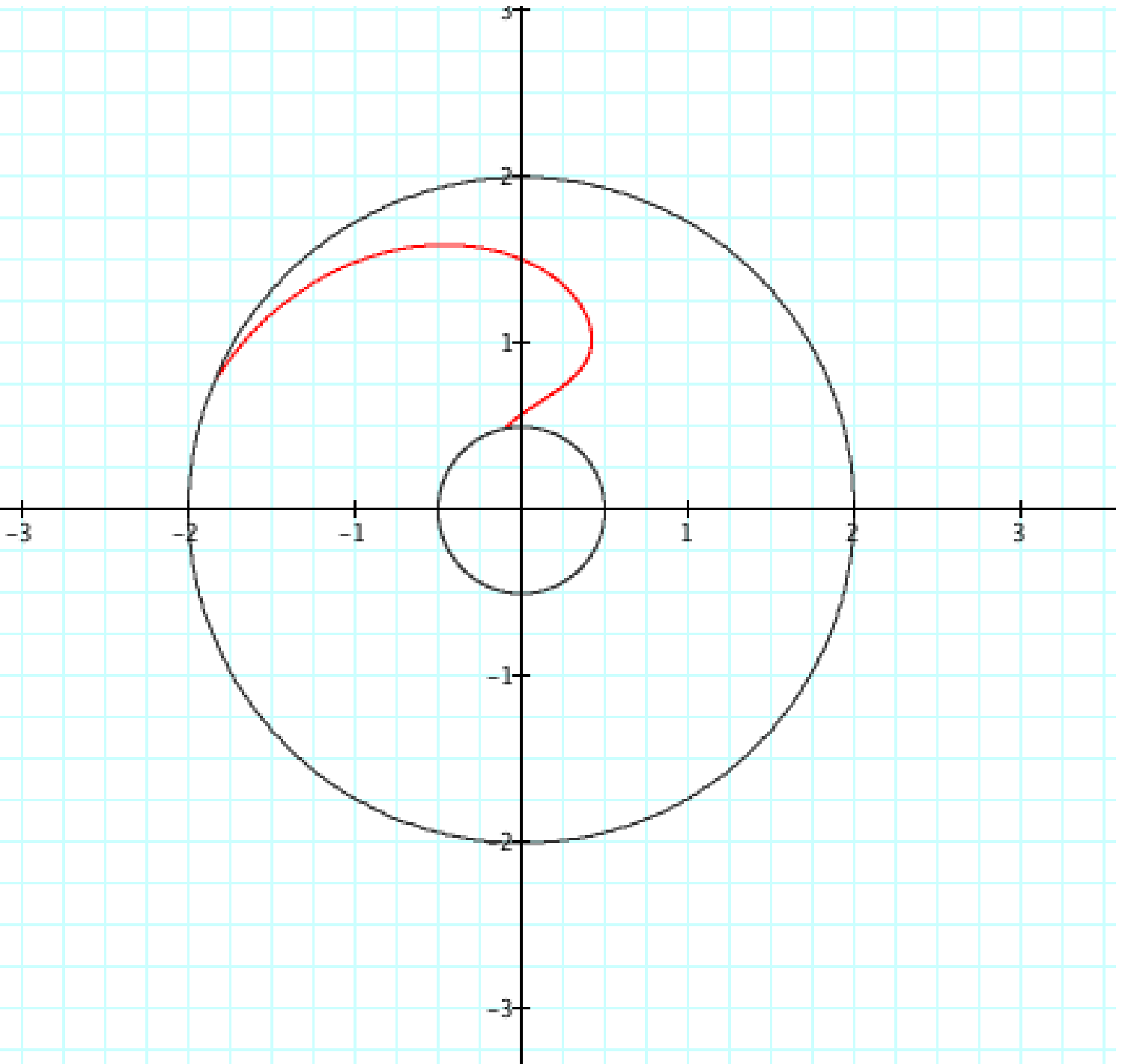}}
\vskip0cm

\parindent=1pc
\b

\hskip 1cmFig.5.10. Snapshot. Contra-rotation. \hskip2.5cm Fig.5.11. Co-rotation.
\b

\ce{\bf Surface tension}

The success of the no-slip boundary condition is witness to a considerable surface tension, to a strong adhesion of the fluid to the walls.  Directly, this affects only the first layer of molecules near the walls, but water has
a great tensile strength, as is known from the ease at which it  withstands negative pressure. See Fig 5.10.

\epsfxsize.5
\hsize
\centerline{\epsfbox{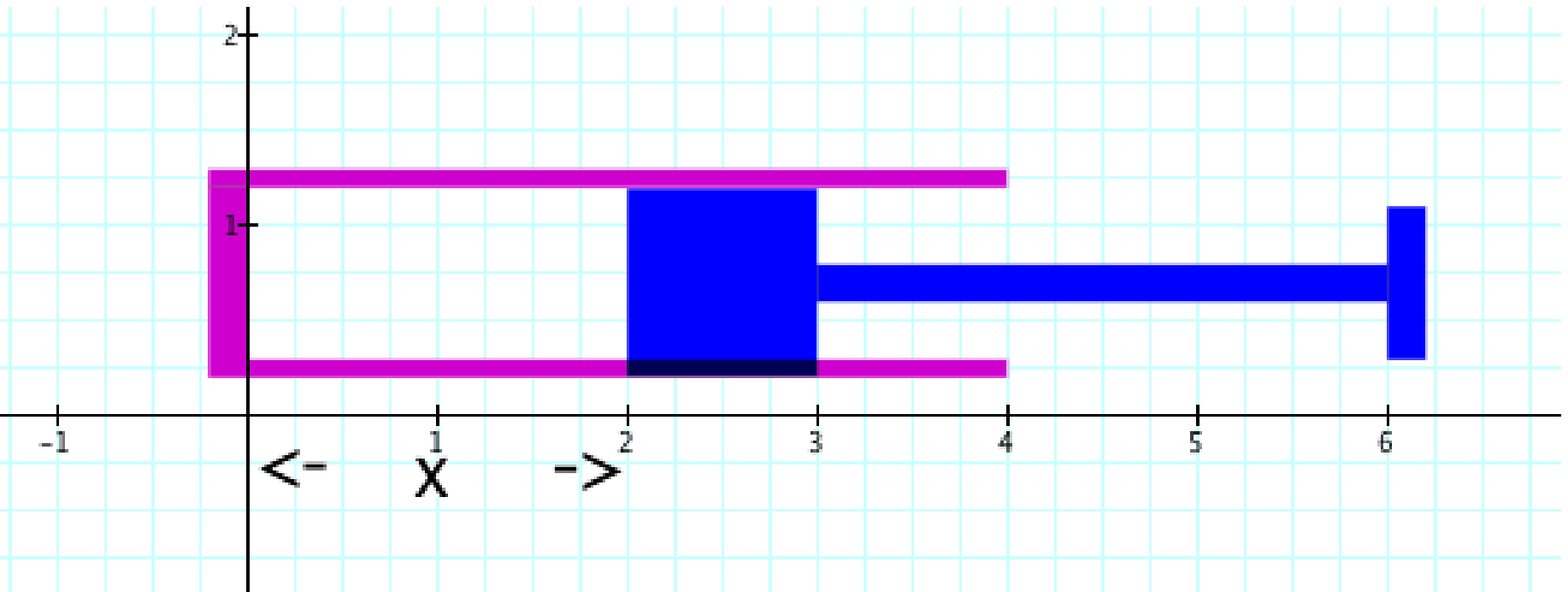}}
\vskip0cm

\parindent=1pc

Fig. 5.12. Negative pressure. 
 The cylinder is filled with water, at atmospheric pressure. The piston is pulled, increasing $x$ and the volume, and the pressure, until the bubble point is reached and the water begins to boil. If care is exercised the system may go into a metastable state and the pressure eventually becomes negative, evidence of the tensile strength of water.  But  responding to any perturbation,  if the laws of adiabatic thermodynamics are obeyed, the water will begin to boil, creating steam with a higher specific entropy density. (It appears as if 
 water stretches like a rubber band, with an increase in entropy. When stretched to the breaking point it evaporates explosively since it already has the entropy of steam; there is no need for further energy transfer. This remark is, of course, highly speculative.)

\b
At the beginning of the experiment the fluid is at rest and the  density is uniform. There is no pressure gradient and no centrifugal force. Then, as we begin to rotate the cylinders the adhesion of the fluid to the wall exerts a force within the fluid. By the principle of Le Chatelier, the reaction of the system is always to resist; hence, at least at first, the pressure and the centrifugal effect must cooperate against the effect of the adhesion. The non-slip boundary condition is not just a mathematical rule; it speaks for the action of a third force. In our model, the kinematic potential and the pressure continue to act together in most of the stable configurations. Perhaps Fig.s
5.11 and 5.12 may help to bring this interpretation in line with intuition.
\b

Consider two volume elements, one on the inner cylinder and the other at the outer cylinder, and a thin smooth tube connecting the two points. Make this more precise by assuming that the tensile strength of the fluid is transmitted along the tube. Then there is one case when this force is constant, when the 
instantaneous, linear velocity of the two end-points are equal. In that case $\omega_i/\omega_o = r_o/r_i$. At that point the centrifugal force has to cancel the force produced by the gradient of the pressure; the pressure increases outwards.

Perhaps that is what we see in Fig. 5.9 when we notice that $K'$ turns positive in a narrow, radial area on the right of the diagram.

\b

\b
\ce{\bf Entropic forces}

The Lagrangian (1.5), in the full thermodynamic setting, is
$$
\L = \rho(\dot\Phi +\dot{\vec X}^2/2 +\kappa\rho \dot{\vec X}\cdot\DD\Phi - \vec\Phi^2/2 - \phi) - f-sT,
$$
where $f(\rho,T)$ is the free energy density. The Bernoulli equation becomes
$$
\DD\big(\dot\Phi +\dot{\vec X}^2/2 +\kappa\rho \dot{\vec X}\cdot\DD\Phi - \vec\Phi^2/2 -\phi\big) = \DD p - (s-\rho{\p s\over \p\rho})\DD T 
-\rho T \DD {\p s\over \p \rho}.  
$$
\no With the usual assumptions, $s = \rho S$, the specific entropy $S$ uniform,
both terms after $\DD p$ drop out. The result is the familiar hydrodynamic
equation; making use of it implies that the specific entropy is assumed to be uniform. The two  additional terms  have been systematically ignored until now, in obedience to a popular and successful strategy:  If an entropy gradient is not needed, ignore it.

In case of difficulties, one may make a different assumption about the entropy density. 
Allowing the  specific entropy $S$ to be non-uniform gives life to the last
term. To invigorate the $\DD T$ term it is necessary to expand the expression $s=\rho S$ with a term that is not linear in $\rho$, as in  Landau and Lifshitz (1956).     

In the present case we have followed the usual protocol and it has not led to any difficulties, the experimental data having been  explained to
surprising accuracy given the speculative nature of (3.3). This may be less  due to a correct theory than to the paucity of experimental data, which may change when the experiment is repeated with a range of different fluids. The system is not at equilibrium; some energy is being supplied; there is likely to be a temperature gradient and there may be some heat flow. Our model calculations depend only on the density and the chemical potential, only 
experiments can determine the pressure and the entropy distribution.
See caption to Fig. 5.10. 
\ve

\no{\bf VI. Conclusions}

Energy, with associated conservation laws, is the very soul of theoretical
physics. It is entirely natural that, in the many contexts where an expression for it is not
available,  attempts are made to invoke it anyway. But our efforts are better spent if we strive to formulate all of physics in terms of action principles. The difficulty, of course, is to determine the action or, before that, to find the variables that make it possible.

The quest for action principles for hydrodynamics and thermodynamics was pursued most vigorously during the late 19'th and early 20'th century, by some of the most important physicists of that pioneering age: Helmholtz, Maxwell, Cartan, Einstein and others. 
In  a memorandum to O. Veblen, dated March 26, 1945, John von Neuman (1945) laments the fact that ``hydrodynamical problems, which ought to be considered relatively simple, offer altogether disproportionate difficulties"; he says  that ``the true technical reason appears to be that variational methods have ... hardly been introduced in hydrodynamics." And he adds:
``It is well known that they could be introduced, but what I would like
to stress is that they have not been used to any practically important
scale for calculations in that field".  
 
The reason that they had not been used is to some extent explained by the internal inconsistencies related to the use (or misuse) of the energy concept.

A viable action principle for rotational Hydrodynamics, Thermodynamics and General Relativity has been proposed (Fronsdal 2014, 2017). It is not unique, but it is the most economical; complicated systems require more variables. This paper
offers a first instance of an application, to a system that has resisted a
complete development even in the special case of incompressible flows.
An application to planetary systems, in the Newtonian approximation,  is in the works and the lifting of this model to General Relativity needs only dedication.

\b

\ce {\bf Future experiments}

In the present contexts of elementary hydrodynamic systems much work has been done on incompressible flows. The more challenging problem of compressible flows require measurements of temperature and pressure profiles, which are difficult. We need more measurements of this type and we need experiments
with wider horizons. Andereck and others have explored the full range of flow regimes in the Couette problem, we need to vary the fluid (even as far as using gases) as well as the density of absorbed air, ambient temperature and pressure and, especially, surface tension; that is, the surface of the cylinders. 
The theory presented here {\bf predicts}  that the slope of the left asymptote
in Fig. 5.1 is sensitive to the compressibility of the fluid.

\b

\ve

\no{\bf References}

\noindent Andereck, C.D., Dickman, R.  and Swinney, H.L.``New flows in a circular Couette 
system

with co-rotating cylinders", Physics of Fluids, {\bf 26}, 1395 (1983); doi 10.1063/1.864328.

\noindent Andereck, C.D., Liu, S.S. and Swinney, H.L.``Flow regimes in a circular Couette 

system with independently rotating cylinders". J.Fluid Mech. {\bf 164} 
155-183  (1986).   

\no Arnold, V.I. and Khesin, B.A., {\it Topological methods in hydrodynamics},
Springer (1998) 

\no Bernoulli,D., {\it Argentorat}, 1738.

\no Betchov, R. and Criminale, W. O. J., Stability of parallel flows, New York, Academic 

Press, 1967.
 
\no Chandrasekhar, S., {\it Hydrodynamics and hydrodynamic stability},
Oxford Univ. Press (1961). 

\noindent Couette, M., ``Oscillations tournantes d'un solide
de r\'evolution en contact avec un fluide

visqueux," Compt. Rend. Acad. Sci. Paris
{\bf 105}, 1064-1067 (1887).

\noindent Couette, M., ``Sur un nouvel appareil pour
l'\'etude du frottement des fluides", 

Compt.Rend. Acad. Sci. Paris {\bf 107}, 388-390 (1888).

\noindent Couette M., ``La viscosit\'e des liquides," 

Bulletin
des Sciences Physiques {\bf 4}, 40-62, 123-133, 262-278 (1888). 

\noindent Couette, M., ``Distinction de deux r\'egimes dans
le mouvement des fluides," 

Journal de
Physique [Ser. 2] IX, 414-424 (1890)

\noindent Couette, M., ``Etudes sur le frottement des
liquides," 

Ann. de Chim. et Phys. [Ser. 6] 21, 433?510 (1890).

\no Drazin, P.G. and Reid, W.H., , {\it Hydrodynamic Stability}, Cambridge University Press (1981)

\no Eddington, A.S., {\it The internal constitution of stars}, Dover, N.Y.
(1959)
 



\no Fetter, A.L. and Walecka, J.D., {\it Theoretical Mechanics of Particles
and Continua}, 

MacGraw-Hill NY 1980.

\noindent Fetter, A.L., ``Rotating trapped Bose-Einstein condensates",
Rev.Mod.Phys. {\bf 81} 647-691 

 (2009).

\no Fronsdal, C. ``Ideal stars in General Relativity", Gen. Rel. Grav {\bf 39}, 1971-2000 (2007) 

\no Fronsdal, C. ``Adiabatic Thermodynamics of Fluids", monograph in progress, fronsdal@physics.ucla.edu.

\noindent Fronsdal, C. , ``Heat and Gravitation. The action Principle", 

Entropy 16(3),1515-1546 (2011).

\noindent Fronsdal, C., ``Action Principle for Hydrodynamics and Thermodynamics 

including general, rotational flows",   arXiv 1405.7138v3[physics.gen-ph], (2014).


\noindent Fronsdal, C., ``Relativistic Thermodynamics, a Lagrangian Field Theory
  for general flows 
  
  including rotation", Int.J. Geom.Math.Meth. Phys., to appear
  (2016a). 
  
   \noindent  Fronsdal, C., ``Action Principles for Thermodynamics, Review and Prospects",  

Phys. Elem. Part. and Nucl.  (JINR) (2017).
  
 \no  Fronsdal, C., ``Action Principles for Thermodynamics"  in   
{\it Memorial Volume on Abdus 

Salam's 90th Birthday}, Lars Brink, Michael Duff and Kok Khoo Phua (eds.), 

 (World Scientific Publishing), ISBN: 978-981-3144-86-6 (2017).


 
 \noindent Hall, H.E. and Vinen, W.F., ``The Rotation of Liquid Helium II. The Theory of
 
 Mutual Friction in Uniformly Rotating Helium II", 

Proc. R. Soc. Lond. A 1956 238, doi: 10.1098/rspa.1956.0215 (1956)

\no \no Khalatnikov, I. M.., {\it an Introduction to the theory of superfluidity}, Benjamin New York 

1965.

\no Kippenhahn, R.,  Weigert, A and Wess, A, {\it Stellar Structure and Evolution}, Springer  2012. 
  
 \no Koschmieder, E.L., {\it Benard cells and Taylor vortices}, Cambridge 1993.
 
 \no Lamb, H. ``Hydrodynamics", Cambridge U. Press 1932.

\no \no  Joseph, D. D., Stability of fluid motions, Vol.1 and 2, Berlin, Springer-Verlag, 1976.

\no Lagrange, J.M.,  Taurinensia, ii., (1760), Oeuvres, Paris, 1867-92.

\no Lagrange, J.M.,  " Memoire sur la ThSorie du Mouvement des Fluides ", Notiv. mem. de 

VAcad. de Berlin, 1781, Oeuvres, t. iv.



\noindent  Lin, C. C., The theory of hydrodynamic stability, Cambridge, 

Cambridge Press, 1955.

\no \noindent Mallock, A., Proc.R.Soc. {\bf 45} 126 (1888).

\noindent Mallock, A., Ohilos.Trans.R.Soc. {\bf 187}, 41 (1896).

 \noindent Ogievetskij, V.I.and I.V.Polubarinov, I.V., ``Minimal interactions between spin 0 and 
 
 spin 1 fields", J. Exptl. Theor. Phys. (USSR) {\bf} {\bf 46}, 1048–1055 (1964).  
 
\noindent Putterman, S.J., ``Superfluid hydrodynamics", North-Holland , Amsterdam 

\no Rayleigh, L., On the stability or instability of certain fluid motions, Proc. Lond. Maths. 

Soc. {\bf 11}, 1880, 57-70. 

\no Rayleigh, L., ``On the dynamics of revolving fluids", Proc.Roy.Soc. A{\bf 93}148-154 (1916).

\noindent Schmid, P.J. and Henningson, D. S., Stability and transition in shear flows, New York, 

Springer-Verlag, 2000.

\no  Taylor, G.I.,  ``Stability of a viscous fluid contained between two rotating cylinders", 

Phil.Trans. R.Soc. London {\bf A102}, 644-667 (1923).

\no Tilley, D.R. and Tilley, J, {\it Superfluidity and Superconductivity}, Adam Hilger Ltd., Bristol 

(1974)  
 
\no Zheltukin, A., ``On brane symmetry, arXiv.1409.6655.

\end

\vskip1cm

\epsfxsize.5
\hsize
\hskip-3cm
\centerline{\epsfbox{CouetteFlow1.eps}}
\vskip0cm

\parindent=1pc\hskip1mm
\b

\vskip-2.3in\hskip1.5in
\epsfxsize.4\hsize
\centerline{\epsfbox{CouetteFlow2.eps}}
\vskip0cm

\parindent=1pc
\b

\hskip 1cmFig.5.11. Snapshot. Contra-rotation. \hskip2.5cm Fig.5.12. Co-rotation.

\no{\bf Appendix: A theory of incompressible fluids}

We set
$$
{\rho_i\over \rho_o} = 1+\xi,~~~~ c = 1 - \psi,~~~0< \psi < \xi
$$
and obtain, on the asymptotes  $K=0$, in the first quadrant, 
$$
{\omega_i\over \omega_o} = ({r_o\over r_i})^2{2-\psi\over 2-\psi+\xi} 
\approx 1.28
$$
and in the fourth quadrant,
$$
{\omega_i/\omega_o} = ({r_o\over r_i})^2{\psi\over \psi - \xi} = 
-1.28{1\over  {\xi\over \psi}-1} < - 1.28.
$$
In the best fit $\xi/\psi \approx 5$.

We get a model of an incompressible gas by taking the limit of large $\kappa$.
In this case it is better to renormalize the field $\vec X$ by absorbing a factor $\kappa$, to express the  the Lagrangian density (1.5) as
$$
\L = \rho(\dot\Phi +\dot{\vec X}^2/2\kappa^2 +\rho \dot{\vec X}\cdot\DD\Phi - \vec\Phi^2/2 - \phi) - W[\rho].\eqno(6.1)
$$
The second term may perhaps be neglected in some applictions.

 The important fields are the velocity of mass flow
$$
\vec v = \dot{\vec X} -\DD\Phi,  
$$
and the renormalized  vector field 
$$
{\vec w\over \kappa} = {1\over \kappa^2}\dot{\vec X}+\DD\Phi.
$$
In the limit $1/\kappa \rightarrow 0$ the field  $\vec X$ becomes a non-dynamic field, as in the approach initiated by Hall and Vinen {\it op cit}. 
\b


 \hskip2cm
\epsfxsize.3\hsize
\centerline{\epsfbox{10.5.1.eps}}
\vskip-.5cm

\parindent=1pc
\bb
\vskip-2cm
Fig. 5.1. The curves are $K$ loci.
\b

\no{\bf 5. Special Relativity. The Notoph gauge theory}

As with all non-relativistic vector fields the problem of a Lorentz covariant 
generalization is urgent. The idea that every 3-vector, non-relativistic field should grow a fourth component has been popular and in particle mechanics one replaces
$$
\dot X^i\rightarrow\left({d\over d\tau} \vec X, X^0\right),~~~X^0 = 
\sqrt{c^2 + \vec X^2}.
$$
But the proper time $\tau$ has no place in a theory of fields and, besides, 
the relativistic vector field contains 3 pairs of presumptive canonically conjugate variables. (Two if massless.)  Instead, the relativistic 2-form has just one propagating mode.

The relativistic, antisymmetric tensor field ($Y_{\mu\nu}$) was first studied by Ogievetskij and Palubarinov (1964). One of its remarkable properties is that it mixes with the electromagnetic field to massify it. Its role in string theory was discovered by Lund and Regge and by Kalb and Ramond ( and others.

The relativistic notoph Lagrangian includes the Maxwell Lagrangian and a mixing term
$$
\L_{OP} = dY^2 + {1\over8\pi}F^2 +\gamma  Y F.\eqno(6.1)
$$
We must include an interaction with the fields $\rho$ and $\Phi$; the most 
natural way is to set the total Lagrangian matter density to be
$$
\L_{\rm matter} = \L_{FW} +{\rho\over 2}dY^2 + + {1\over8\pi}F^2 +\gamma  Y F +{\kappa \rho\over \sqrt{-g}} d\psi\w dY.\eqno(6.2)
$$
The last term is included since it is gauge invariant, but the electromagnetic terms will be neglected in the present context. The coefficients $\gamma$ and $\kappa$ are constants.

The ``notoph''  field has six components, 
$$
Y_{ij} = \epsilon_{ijk}X^k, ~~~~Y_{0i} = \eta_i, ~~~i,j = 1,2,3.
$$
The gauge group consists of the transformations
$$
\delta Y = d\w\xi,
$$
with an arbitrary one-form $\xi$;
by this means the field $\vec \eta$ can be reduced to zero, leaving the field
$\dot{\vec X}$ that has been associated with fluid flow since antiquity.

Written out in full, our present relativistic  Lagrangian density is
$$
\L = {\rho\over 2}(g^{\mu\nu}\psi_{,\mu}\psi_{,\nu}-c^2) + {\rho\over 2} dY^2  +\kappa\rho{c^2\over 2}\epsilon^{\mu\nu\lambda\rho}Y_{\mu\nu,\lambda}\psi_{,\rho}  -f -sT,\eqno(6.3)
$$

See Ogievetski and Palubarinov (1964), Kalb and Ramond (), Lund and Regge (),
Fronsdal, (2011).
\b

 \no{\bf 7.  Non relativistic limit and Galilei transformations}

A concept of a non-relativistic limit of a relativistic field theory can be
envisaged if each of the dynamical variables can be represented as a power series
in $1/c$, beginning with a term of order zero; that is, $(1/c)^0$, or higher,
$(1/c)^1,  (1/c)^2, ... $~. In the case of the model considered we must assume that this is the case for the variables $\rho,Y, \psi$. We neglect the coupling to the electromagnetic field for the moment. The non-relativistic limit of the Lagrangian
exists if every term is of positive or zero order. Dropping all terms of positive order we may ask about the physical meaning  of the remainder, including transformation properties.

Taking the basic variables to be $\rho, Y$ and $\psi$ we find, with the normalization
$$
dY_{\mu\nu\lambda} = \sum_{cyclic}Y_{\mu\nu,\lambda},\eqno(7.1)
$$
that 
$$
{c^2\over 24}dY^2 = {1\over 2}(\dot{\vec X} +\DD\w\eta)^2
 - {c^2\over 4}(\DD\cdot\vec X)^2,\eqno(7.2)
 $$
 $$
{\kappa\over 2}\epsilon^{\mu\nu\lambda\rho}Y_{\mu\nu,\lambda}\psi_{,\rho}
 = \kappa(\dot{\vec X}+\DD\w\eta)\cdot\DD\Phi -\kappa c^2\DD\cdot \vec X
 $$
There are terms of order $c^2$ and to overcome this obstruction we must postulate the boundary condition
 $$
 \DD\cdot \vec X = c^{-2}   \Theta + o(c^{-3}),\eqno(7.3)
 $$
with the field  $\Theta$ of order 0. 
We conclude that the existence of a non-relativistic limit of the Lagrangian depends on the validity of (7.1).
\bb

\ce{\bf Galilei invariance}

 The subgroup of `proper' Galilei transformations derives from Lorentz transformations of the form
$$
\delta\vec x = t\vec u\gamma,~~~~\delta t = (\vec u\cdot \vec x/c^2)\gamma,~~~~\gamma = {1\over \sqrt{1-(u/c)^2}}.
$$
Infinitesimal Galilei transformations are related to first order Lorentz transformations. It is enough to retain terms linear in $\vec u$, 
replacing $\gamma$ by unity. Infinitesimal Galilei transformations are defined
as the `contraction' that consists of taking the limit $c r \infty$.But it would be imprudent to take that limit already at this stage, as we shall see. So the transformations to be considered are first order or infinitesimal  Lorentz transformations, 
$$
\delta\vec x = t\vec u,~~~~\delta t = \vec u\cdot \vec x/c^2.  
$$
 
In what we shall call a physical gauge the field $\vec \eta$ vanishes. The
Lorentz group acts on the antisymmetric field in the manner that is indicated by the indices, in particular
$$
\delta Y_{ij} = t\vec u\cdot\DD Y_{ij}+ tu_i Y_{0j} + tu_jY_{i0},~~~~
 \delta Y_{0j} = t\vec u\cdot\DD Y_{0j} +  u_iY_{ij}/c^2.
$$
The terms that involve the operator $\vec u\cdot\DD$ are generic for scalar fields and they do not contribute to the action.
\b

\ce{\bf The energy-momentum tensor}

The energy momentum tensor is based on Cauchy's stress tensor, but the basic role that it plays in relativistic field theories was explained by Emmy Noether (191 ).
In any special-relativistic field theory with Lagrangian density $\L$ it is the result of evaluating the integrals
$$
\delta \int d^4x \L = \int d^4x G\L,
$$
where $G$ is a generator of infinitesimal translations, on shell. . If $\L$ is expressed entirely in terms of a family of fields $\alpha,...$ and their first order derivatives $\alpha_\mu,...$, then the variation is
$$
\int d^4x G\L = \sum_\alpha\int d^4x\left({\p\L\over \p\alpha}G\alpha
+ \sum_\mu{\p\L\over \p\alpha_{\mu}}\p_\mu G\alpha\right).
$$
An integration by parts of the second term, and use of the Euler-Lagrange equations, leaves only the `integrated parts'
$$
\int d^4x G\L = \int d^4x \sum_\mu\p_\mu\left(\sum_\alpha{\p\L\over \p\alpha_{\mu}} G\alpha\right)\eqno(3.4)
$$
Taking $G = \p_\nu,~ \nu = 0,1,2,3,$ this becomes
$$
\p_\mu T_\nu^\mu = 0,~~~
T_\nu^\mu = \sum_{\alpha := \Phi,\rho}{\p\L\over \p\alpha_{\mu}}\alpha_{\nu}
-\delta_\nu^\mu\L,\eqno(3.5)
$$

The Navier-Stokes equation was deduced from properties of a phenomenological  stress tensor, but the dynamics behind the energy momentum tensor was explained much later (Noether 191 ).The importance of the energy-momentum tensor comes from its relation to conservation laws, but an action principle is needed to imbed the conservation laws in a coherent, dynamical theory.

In the special case of potential flow the action of Fetter and Walecka gives the energy momentum tensor with components
$$
T_0^0 = \rho\dot\Phi -\L = h,~~~{\rm energy~density},\eqno(3.6)
$$
$$
T_0^i =   -\dot\Phi\rho \Phi_{,i},~~~{\rm energy~flow~density}\eqno(3.7)
$$
$$
T_i^0 = \rho\Phi_{,i}, ~~~{\rm momentum~density}\eqno(3.8)
$$
and stress tensor
$$
T_i^j=  -\Phi_{,i} \rho  \Phi_{,j} - \delta_i^j\L. 
$$
We can verify the conservation law (3.5). It splits into these two cases. the first one,
$$
\p_0T_0^0 + \p_jT_0^j = 0\eqno(3.9)
$$
reduces to
$$
{d \over dt} \, h + \DD\cdot(\rho \dot\Phi\vec v) = 0,~~~v_i := -\Phi_{,i},
$$
or
$$
{\p\over \p t}h + \DD\cdot (h+p)\vec v = 0.\eqno(3.10)
$$
When this is integrated over a domain $\Sigma$ the second term reduces to a surface integral over the boundary $\p\Sigma$, which vanishes by virtue of the boundary conditions if $\Sigma$ is the extent of the vessel that contains the fluid. Also, if $\Sigma$ is any domain that moves with the fluid, so that the flow on the boundary $\p\Sigma$ is tangential, then this term makes no contribution. Hence, in both these cases, the total energy in $\Sigma$ is a constant of the motion.
This justifies referring to (4.6) as a statement of energy conservation. 
 The other one,
$$ 
\p_0T_i^0+\p_jT_i^j = 0
$$
becomes
$$
{d\over dt}(\rho v_i) + \p_j(\rho v_j v_i) = - \p_ip,~~~i = 1,2,3;\eqno(3.11)
$$
it is often referred to as a statement of momentum conservation, which is
somewhat misleading. If (3.8) expresses a conservation law it can only be the conservation of $\rho\vec v$.  Using the equation of continuity we can rewrite it as follows
$$
{d\over dt} v_i + (\vec v\cdot\DD) v_i =-{1\over \rho}\p_ip,~~~i = 1,2,3,\eqno(3.12)
$$
which agrees with the Bernoulli equation.  Eq.(3.12) is the Navier - Stokes equation for this case.

Let us consider the case $G= \vec x\w \DD$; then instead of (4.12)
$$
{d\over dt} \vec x\w\vec v + (\vec v\cdot\DD) \vec x\w\vec v =-{1\over \rho}\vec x\w \DD p,~~~i = 1,2,3,\eqno(3.13)
$$
in the case of cylindrical Couette flow the vertical component of the right hand side and the vertical component of the angular momentum are conserved
in a fluid element that moves with the flow.  The analysis can be extended to include the contributions of the field $\vec X$, without any instructional gain.
 \b

\ce{\bf Boundary conditions}

The walls of the two cylinders move with angular  velocities

$$
\omega_i\hat\theta = {\omega_i\over r}(-y,x,0),~~ \omega_o\hat\theta
 = {\omega_o\over r}(-y,x,0)~~~\omega_i,  \omega_o~~{\rm constant}.
$$
The velocity of mass transport is $\vec v$, Eq.(4.4), this is the velocity that must satisfy  the no-slip boundary conditions, whence
$$
{1+\kappa^2\over r_i^2}(a+{bc^2\over\rho_i}) = \omega_i,
~~~{1+\kappa^2\over r_0^2}(a+{bc^2\over\rho_o}) = \omega_o,\eqno(4.10)
$$
and
$$ 
a = {r_o^2\rho_o\omega_o - r_i^2\rho_i\omega_i\over (\rho_o-\rho_i)(\kappa^2+1)},
~~~~ b = \rho_o\rho_i{r_i^2\omega_i - r_o^2\omega_o\over \kappa^2(\rho_o-\rho_i)}.\eqno(4.11)
$$

At the outer boundary $r=r_o=1$ and $\rho = \rho_o = 1$, and  (4.7-8) reduce to
$$
K = {1+\kappa^2\over 2}(b^2c^2-a^2),~~~~K' = 2\big(a^2 -(1+2\alpha)b^2c^2\big).
\eqno(4.12) 
$$

The alternative assumption, that density is highest at the outer boundary
and that, as the speed is increased,the instability manifests itself there,   is not successful. At the outer boundary (5.1-2) are replaced by  
$$
{\omega_i\over \omega_0} = ({r_o\over r_i})^2{1\pm c(\rho_o/\rho_i)\over 1\pm c}. 
$$
and $1> c > {\rho_i\over \rho_o}. $
 
   The calculation gives loci just like that in Fig. (5.3), the best fits with $\rho_i = .9, \kappa = 2.27,
\rho_i = .99, \kappa = 7.7, \rho_i = .999, \kappa = 24.4$. see Fig.5.6. The six points lie on the parabolas $x = -.167 y^2$ and $x=.33 y^2$.

However, the system does not behave as expected in this case, so this alternative can be ignored, especially since the assumption that $\rho_i>1$
works exceedingly well.

\epsfxsize.4\hsize
\centerline{\epsfbox{Rayleigh4.6.eps}}
\vskip0cm

\parindent=1pc
 
Fig.5.4. The relation between $\rho_i/\rho_o$ and $\kappa$ that assures a best fit  to experiments, in the case that the instability manifests itself at the outer (inner) boundary on the left (right). The abscissa is $1/(\rho_i-1)$, the ordinate is $\kappa$.
  
\epsfxsize.4\hsize
\vskip0cm

\parindent=1pc

\end